\documentclass[useAMS,usenatbib]{mn2e}

\usepackage{psfig}
\usepackage{graphicx}
\usepackage{times}
\usepackage{natbib}

\newif\ifAMStwofonts
\AMStwofontstrue

\newcommand{\be}{\begin{equation}}
\newcommand{\ee}{\end{equation}}
\newcommand{\ba}{\begin{eqnarray}}
\newcommand{\ea}{\end{eqnarray}}
\newcommand{\brr}{\begin{array}}
\newcommand{\err}{\end{array}}
\newcommand{\bc}{\begin{center}}
\newcommand{\ec}{\end{center}}
\newcommand{\hm}{\,h^{-1}{\rm Mpc}}

\newcommand{\msun}{\,h^{-1}M_\odot}

\newcommand{\mincir}{\raise
  -2.truept\hbox{\rlap{\hbox{$\sim$}}\raise5.truept \hbox{$<$}\ }}
\newcommand{\magcir}{\raise
  -2.truept\hbox{\rlap{\hbox{$\sim$}}\raise5.truept \hbox{$>$}\ }}
\newcommand{\siml}{\raise
  -2.truept\hbox{\rlap{\hbox{$\sim$}}\raise5.truept \hbox{$<$}\ }}
\newcommand{\simg}{\raise
  -2.truept\hbox{\rlap{\hbox{$\sim$}}\raise5.truept \hbox{$>$}\ }}

\title[Joint deprojection of SZ and X--ray images] 
{Joint deprojection of Sunyaev--Zeldovich and X--ray images of galaxy
clusters} \author[Ameglio et al.]  {S. Ameglio$^{1,2,3}$,
S. Borgani$^{1,2,3}$, E. Pierpaoli$^4$ \& K. Dolag$^5$ \\~\\ $^1$ Dipartimento di
Astronomia dell'Universit\`a di Trieste, via Tiepolo 11, I-34131
Trieste, Italy (ameglio,borgani@ts.astro.it) \\ 
$^2$ INFN -- National Institute for Nuclear Physics, Trieste, Italy\\ 
$^3$ INAF -- National Institute for Astrophysics, Trieste, Italy\\ 
$^4$ University of Southern California, Los Angeles, CA
(pierpaol@usc.edu)\\  
$^5$ Max-Planck-Institut f\"ur Astrophysik, Karl-Schwarzschild Strasse
  1, Garching bei M\"unchen, Germany (kdolag@mpa-garching.mpg.de)}

\begin{document}

\date{Accepted ???. Received ???; in original form ???}

\maketitle

\begin{abstract}
  We present two non--parametric deprojection methods aimed at
  recovering the three-dimensional density and temperature profiles of
  galaxy clusters from spatially resolved thermal Sunyaev--Zeldovich
  (tSZ) and X--ray surface brightness maps, thus avoiding the use of
  X--ray spectroscopic data. In both methods, the cluster is assumed
  spherically symmetric and modeled with an onion--skin structure.
  The first method follows a direct geometrical approach, in which the
  deprojection is performed independently for the tSZ and X--ray
  images, and the resulting profiles are then combined in order to
  extract density and temperature.  The second method is based on the
  maximization of a single joint (tSZ and X--ray) likelihood
  function. This allows to fit simultaneously the two signals by
  following a Monte Carlo Markov Chain approach.  These techniques are
  tested against both an idealized spherical $\beta$--model cluster
  and against a set of clusters extracted from cosmological
  hydrodynamical simulations with and without instrumental noise. In
  the first case, the quality of reconstruction is excellent and
  demonstrates that such methods do not suffer of any intrinsic
  bias. As for the application to simulations, we projected each
  cluster along the three orthogonal directions defined by the
  principal axes of the momentum of inertia tensor. This enable us to
  check any bias in the deprojection associated to the cluster
  elongation along the line of sight. After averaging over all the
  three projection directions, we find an overall good
  reconstruction, with a small ($\mincir 10$ per cent) overestimate of
  the gas density profile.  This turns into a comparable overestimate
  of the gas mass within the virial radius, which we ascribe to the
  presence of residual gas clumping. A part from this small bias the
  reconstruction has an intrinsic scatter of about 5 per cent, which
  is dominated by gas clumpiness. Cluster elongation along the
  line of sight biases the deprojected temperature profile upwards at
  $r\mincir 0.2r_{\rm vir}$ and downwards at larger radii. A
  comparable bias is also found in the deprojected temperature
  profile. Overall, this turns into a systematic underestimate of the
  gas mass, up to 10 percent.  We point out that our recovered
  temperature profiles are much closer to the mass--weighted profiles
  than those obtained from the X--ray spectroscopic--like
  temperature. These results confirm the potentiality of combining tSZ
  and X--ray imaging observations to the study of the thermal
  structure of the intracluster medium out to large cluster-centric
  distances.
\end{abstract}

\begin{keywords}
large-scale structure of Universe -- galaxies: clusters: general --
cosmology: miscellaneous -- methods: numerical
\end{keywords}

\section{Introduction}
A precise observational characterization of the thermal structure of
the intra--cluster medium (ICM) is of crucial relevance for at least
two reasons. On one hand, the ICM thermodynamics is determined not
only by the gravitational accretion of gas into the dark matter (DM)
potential wells forming clusters, but also by energy feedback
processes (i.e., from supernova explosions and active galactic
nuclei), which took place during the cosmic history of the cluster
assembly. On the other hand, a precise characterization of the
temperature structure of clusters is highly relevant to infer the
cluster masses, under the assumption of hydrostatic equilibrium, and,
therefore, to calibrate clusters as precision tools for cosmological
applications \citep[e.g., ][for
reviews]{2002ARA&A..40..539R,2005RvMP...77..207V,2006astro.ph..5575B}.

The study of the ICM properties has been tackled so far through X--ray
observations. Data from the Chandra and XMM--Newton satellites are
providing precise measures of the temperature and surface brightness
profiles for a fairly large number of nearby ($z\mincir 0.3$)
clusters, reaching $z\simeq 0.5$ for the brightest objects
\citep[e.g.,
][]{2005A&A...433..101P,2005A&A...429..791P,2005ApJ...628..655V,2006ApJ...641..752K}.

These observations have indeed allowed  to trace
in detail the mass distribution in galaxy clusters for the first time.
However   in the X-rays the 
accessible dynamic range is limited by the $\rho_{gas}^2$ dependence
of the emissivity which causes  measurements of the
temperature profiles to  be generally limited to 2--3 core radii,
extending out to $r_{500}$ only in the most favorable cases.
This is not the case for 
clusters' studies
performed with the Sunyaev--Zeldovich effect
(\citealt{1972CoASP...4..173S}, tSZ hereafter; see
\citealt{1999PhR...310...97B,2002ARA&A..40..643C} for reviews). Since the
tSZ signal has a weaker dependence on the local gas density, it is in
principle better suited to sample the outer cluster's regions, which can
be accessed by X--ray telescopes only with long exposures and a
careful characterization of the background noise. Clusters are
currently observed through their thermal SZ (tSZ) signal and
tSZ surveys of fairly large area of the sky promise to discover in the
next future a large number of distant clusters out to $z\magcir 1$.

Thanks to the different dependence of the tSZ and X--ray emission on
the electron number density $n_e$, and temperature $T_e$, the
combination of these two observations offers in principle an
alternative route to X--ray spectroscopy for the study of the
structural properties of the ICM.  Indeed, while the X--ray emissivity
scales as $n_e^2\Lambda(T)$ (where $\Lambda(T)$ is the cooling
function), the tSZ signal is proportional to the gas pressure,
$n_eT_e$, integrated along the line-of-sight. Recovering the
temperature structure of galaxy clusters through the combination of
X--ray and tSZ data has several advantages with respect to the more
traditional X--ray spectroscopy. First of all, surface brightness
profiles can be recovered with a limited number ($\sim 10^3$) of
photons, while temperature profiles require at least ten times more
counts. Therefore, the combination of X--ray surface brightness and
tSZ data should allow to probe more easily the regimes of low X--ray
surface brightness (i.e. external cluster regions and high--redshift
galaxy clusters), which are hardly accessible to spatially resolved
X--ray spectroscopy. Furthermore, fitting X--ray spectra with a single
temperature model is known to provide a temperature estimate which is
generally biased low by the presence of relatively cold clumps
embedded in the hot ICM atmosphere
\citep{2004MNRAS.354...10M,2006ApJ...640..710V}. On the other hand,
combining X--ray and tSZ does not require any spectral fitting
procedure and, therefore, yields a temperature which is basically
mass--weighted.

The combination of X--ray and tSZ observations is currently used to
estimate the angular diameter distance of clusters \citep[e.g., ][,
and references therein]{2006ApJ...647...25B,2006MNRAS.369.1459A} and
to recover the gas mass fraction \citep[e.g.,][]{2006ApJ...652..917L}.
Clearly, performing a spatially--resolved reconstruction the thermal
structure of the ICM requires the availability of high--resolution tSZ
observations with a sub-arcmin beam size, with a sensitivity
of few $\mu$K on the beam. Although observations of this type can not
be easily carried out with millimetric and sub--millimetric telescopes
of the present generation, they are certainly within the reach of
forthcoming and planned instruments of the next generation, based both
on interferometric arrays (ALMA: Atacama Large Millimeter
Array\footnote{\tt http://www.eso.org/projects/alma/}) and on single
dishes with large bolometer arrays (Cornell--Caltech Atacama
Telescope:
CCAT\footnote{http://astrosun2.astro.cornell.edu/research/projects/atacama/};
Large Millimeter Array: LMT\footnote{http://www.lmtgtm.org/}).

Combining X--ray and tSZ data to reconstruct the three dimensional gas
density and temperature structure of galaxy clusters is not a new idea
and different authors have proposed different
approaches. \cite{1998ApJ...500L..87Z} used a deprojection method,
based on Fourier transforming tSZ, X--ray and lensing images, under
the assumption of axial symmetry of the cluster. After applying this
method to simple analytical cluster models, they concluded that the
combination of the three maps allows one to measure independently the
Hubble constant $H_0$ and the inclination angle. This same method was
then applied by \cite{2001ApJ...561..600Z} to cosmological
hydrodynamical simulations of galaxy clusters, who found that a
reliable determination of the cluster baryon fraction, independent of
the inclination angle.  \cite{2000A&A...364..377R} applied a method
based on the Richardson--Lucy deconvolution to combined tSZ, X--ray
and weak lensing data to a set of simulated clusters.
\cite{2001A&A...375...14D} used a perturbative approach to describe
the three dimensional structure of the cluster, to combine tSZ and
lensing images. In this way, they were able to predict the resulting
X--ray surface brightness. After testing their method against
numerical simulations of clusters, they conclude that the DM and gas
distributions can both be recovered quite
precisely. \cite{2004ApJ...601..599L} proposed a method, based on
assuming a polytropic equation of state for gas in hydrostatic
equilibrium, which allowed them to recover the three dimensional
profiles of clusters using the tSZ and the X--ray
signals. \cite{2006A&A...455..791P} applied the same method of
\cite{2000A&A...364..377R} to deproject X--ray and tSZ maps, so as to
recover the gas density and the temperature structure of clusters,
under the assumption of axial symmetry.  \cite{2006ApJ...647L...5C}
applied the combination of tSZ and X--ray observations to recover
determine the ICM entropy profile.

As for applications to real clusters, \cite{2000ApJ...545..141Z}
combined X-ray surface brightness and tSZ data, for a compilation of
clusters, to estimate the central cluster temperature, and found it to
be in reasonable agreement with the X--ray spectroscopic
determination. \cite{2002A&A...387...56P} used ROSAT--HRI imaging data
of a relatively distant cluster ($z\simeq 0.42$) with tSZ
observations to infer the global temperature of the system.

\cite{2005ApJ...625..108D} combined X--ray and tSZ data to constrain
  the intrinsic shapes of a set of 25 clusters.  By applying a
  deprojection method based on assuming the $\beta$--model
  \citep{1976A&A....49..137C}, they confirmed a marginal
  preference for the clusters to be aligned along the line-of-sight,
  thus concluding that X--ray selection may be affected by an
  orientation effect.  \cite{2007arXiv0707.0572S} analyzed the
  potentiality of combining tSZ, X--ray and lensing data to constrain
  the 3D structure of the clusters. He found that these data are
  enough to determine the elongation along the line of sight (together
  with the distance), without however fully constraining shape and
  orientation.

Some of the detailed methods applied to numerical cluster models
account for the presence of a realistic noise in the tSZ and X--ray
maps. However, they generally do not present any detailed assessment
of how this noise determines the uncertainties in the deprojected
profiles, which ultimately characterize the ICM thermodynamics. Having
a good control on the errors is especially crucial in any deprojection
technique, since errors at a given projected separation
affect the deprojected signal in the inner regions, thereby
introducing a non--negligible covariance in the reconstruction of the
three-dimensional profiles.

In this paper we discuss a method to recover the three--dimensional
temperature and gas density profiles from the joint deprojection of
X--ray surface brightness and spatially resolved tSZ data, testing its
performance against idealized spherical clusters and full cosmological
hydrodynamical simulations. This method is based on the assumption of
spherical symmetry, but do not assume any specific model for the gas
density and temperature profiles. We will describe two different
implementations. The first one is analogous to that already applied to
deproject spectroscopic X--ray data \citep[e.g.,
][]{1983ApJ...272..439K} and is based on assuming a onion--like
structure of the cluster, in which projected data of X--ray and tSZ
``fluxes'' are used to recover gas density and temperature in the
external layers and then propagated to the internal layers in a
iterative way. The second implementation is based instead on a
multi--parametric fitting procedure, in which the fitting parameters
are the values of gas density and temperature within different
three--dimensional radial bins. The values of these parameters are
then obtained through a Monte Carlo Markov Chain maximum likelihood
fitting by comparing the resulting projected X--ray and tSZ profiles
to those obtained from the maps. As we shall discuss in detail, this
second method naturally provides the error correlation matrix, which
fully accounts for the covariance between error estimates at different
radii and among different (i.e. gas density and temperature)
profiles. The quality of the X--ray data required by our methods are
basically already available with the current generation of X--ray
telescopes. As for the tSZ data, exploiting the full potentiality of
the deprojection requires spatially resolved data. For illustrative
purposes, we will assume the forecast observing conditions and
sensitivity of the CCAT \citep[][see also
http://www.submm.caltech.edu/$\sim$sradford/ccat/doc/2006-01-ccat-feasibility.pdf]
{2006SPIE.6267E..75S}, although our computations can be easily
repeated for other telescopes.

The plan of the paper is as follows. In Section 2 we describe the
two implementations of the deprojection method, while we describe in
Section 3 their application on a spherical polytropic
$\beta$--model. Section 4 presents the results of the analysis on the
hydrodynamical simulations of clusters. The main conclusions of our
analysis are summarized in Section 5.

\section{The methods of deprojection}

\subsection{The geometrical deprojection technique}
\begin{figure}
\centerline{
\psfig{file=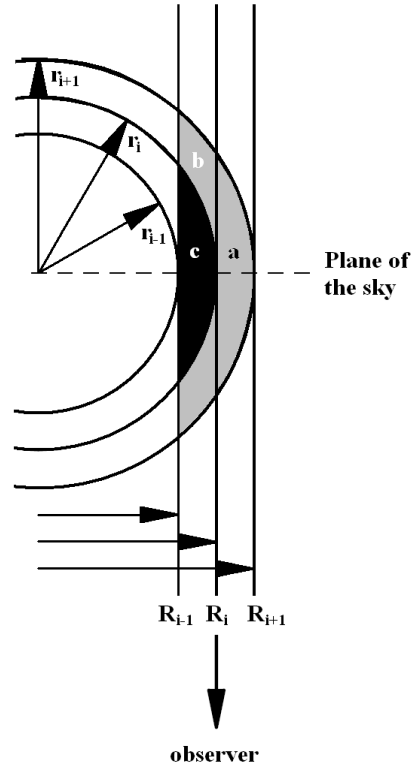,width=6.cm}
}
\caption{Illustration of the onion--skin model adopted for the
geometrical deprojection (see text in Sect. 2.1; adapted from
\citealt{1999AJ....117.2398M}).} 
\label{fi:shells}
\end{figure}

The first method that we apply to recover the three--dimensional
profiles of temperature and gas density is based on a geometrical
technique originally introduced by \cite{1983ApJ...272..439K}, and
subsequently adopted by \citep[e.g.,][]{2000ApJ...539..172B,
2002A&A...391..841E, 2007MNRAS.tmp..541M} to deproject X--ray images
and spectra of galaxy clusters. This method of geometrical
deprojection is fully non--parametric and allows to reconstruct the
3-dimensional profile of a given quantity from its 2-dimensional
observed projection, under the assumption of spherical symmetry.

Following \cite{1983ApJ...272..439K}, the cluster is assumed to have a
onion--like structure (see Figure \ref{fi:shells}), with $N$
concentric spherical shells, each characterized by uniform gas density
and temperature within it. Therefore, the cluster image in projection
is divided into rings, which are generally assumed to have the same
radii of the 3D spherical shells.  Let us define $\epsilon_i$ as the
signal to be recovered from the deprojection method within the $i$-th
shell.  In our analysis $\epsilon_i$ will be proportional to either
$n_eT_e$ for the tSZ signal, or to $n_e^2 \Lambda(T)$ for the X--ray
emissivity. In this way, the contribution of the $i$-th shell to the
surface brightness\footnote{For the sake of clarity, we indicate here
with surface brightness the projected quantity, which can be both a
genuine X--ray surface brightness and the tSZ signal.}  in the ring
$j$ of the image will be given by $s_{i,j}=\epsilon_i \cdot V_{i,j} /
A_j$, where the matrix $V_{i,j}$ has as entries the values of the
volume of the shell $i$ which is projected on the ring $j$, whose area
is $A_j$. By definition, $s_{i,j}=0$ for $j > i$. Accordingly, the
surface brightness $S'_j$ in the ring $j$ can be obtained by summing
up the contributions from all the shells,
\be
\label{eq:deproj}
S'_j = {1 \over A_j} \sum_{i=j}^N s_{i,j} = {1 \over A_j} \sum_{i=j}^N
 \epsilon_i \cdot V_{i,j}\,,
\ee 
where the sums extend over the $N$ radial bins. The deprojection
  amounts to invert the above equation, i.e. to recover the values of
  $\epsilon_i$ from the observed projected signal $S'_j$. We refer to
  Figure \ref{fi:shells} to illustrate how this deprojection is
  performed in practice. Let the shell $i$, limited by $r_{i}$ and
  $r_{i+1}$, be the outermost one. Then, from the surface brightness
  $S'_i$ in the ring $i$ (limited by $R_i$ and $R_{i+1}$), one can
  directly compute the emissivity of the shell $i$ simply by knowing
  the volume of the region (a) and the area of the ring. In this case,
  the sum in eq. \ref{eq:deproj} has only the term $j=i=N$. The
  adjacent inner ring, having index $i-1$ and limited by $R_{i-1}$ and
  $R_{i}$, takes instead a contribution from both the $i-1$ and $i$
  shells. The former is computed by multiplying the emissivity of that
  shell by the volume of the region (b). After subtracting it, the
  only remaining contribution is that of the sell $i-1$ from which the
  emissivity $\epsilon_{i-1}$ is computed. This procedure is then
  repeated from ring to ring down to the center of the cluster.

For this simple scheme to be applied, one requires to have images
extended out to the true external edge of the cluster, i.e. out to the
radius where the surface brightness goes virtually to zero. Clearly,
this situation is never attained in practical applications for at
least two reasons. First, clusters are always embedded in a
large--scale cosmic web, which makes it difficult to define a sharp
outer boundary. Second, and more important, both instrumental and
cosmic backgrounds often dominate the genuine signal from the cluster
well before its virial boundary is reached.

To overcome this problem, it is then necessary to take into account
the emission from the gas, which extends outside the $N$-th
shell. This emission does not have a corresponding ring in the image
but can give a non--negligible contribution to the surface brightness
in all rings. To account for this contribution, we follow the approach
of \cite{1999AJ....117.2398M}, who modeled the volume emission from
the gas beyond the last observable annulus as a power law,
$\epsilon(r) \propto r^{-\alpha}$ (we refer to their Appendix A for a
more detailed description). The idea behind this method is that the
exact contribution to each ring from the external part can be
calculated by integrating the volume emission $\epsilon (r)$.  Then,
the normalization of the power law shape of $\epsilon(r)$ is fixed by
the requirement of matching the total surface brightness of the last
ring. This correction can be expressed as an additional term to
eq.(\ref{eq:deproj}), which is proportional to the surface brightness
of the last ring: \be\label{eq:deproj_edge}
S_j = S'_j + f_j \cdot S_N\,,
\ee Here, $f_j$ is a geometrical factor which is uniquely specified by
the values of the limiting radii of the $j$-th ring and by the
exponent $\alpha$. Eq.(\ref{eq:deproj_edge}) must be actually
interpreted as a set of $2N$ equations, which corresponds to the
separate deprojection of the tSZ and of the X--ray signal, each
performed for $N$ radial bins.  The geometrical deprojection is then
performed by inverting each set of $N$ equations, starting from the
outermost bin and proceeding inward.  This procedure provides the
radial profiles of $n_eT_e$ and of $n_e^2 \Lambda(T)$, whose
combination finally gives the 3D profiles of electron number density
and of temperature. We emphasize that the temperature so obtained is
the actual electron temperature and not the deprojected spectroscopic
temperature, usually obtained from the fitting of X--ray spectra.

Given the iterative nature of this procedure, the uncertainty
associated to each ring propagates not only to the corresponding 3D
shell, but also to all the inner shells.  For this reason, it is very
difficult with this method to have a rigorous derivation of the
statistical uncertainties associated to the deprojected profiles. This
is particularly true for the X--ray profiles, that also  involve  a
derivative of the cooling function with respect to the
temperature. The commonly adopted solution is based on realizing
MonteCarlo simulations, over which to compute the
errors \citep[e.g.][]{2002A&A...391..841E}.

Furthermore, errors associated to different radial bins are not
independent. This is due to the fact that the projected signal in a
given ring is contributed by several shells. The resulting covariance
in the signals recovered in different shells is not provided by this
deprojection method. This is a rather important point on which we will
come back in Section 4.

\subsection{The maximum likelihood deprojection}
This technique is based on performing the deprojection by maximizing a
likelihood function, which is computed by comparing the observed tSZ
and X--ray profiles with the ones obtained by projecting the
onion--skin model in the plane of the sky.  This approach offers more
than one advantage with respect to the geometrical deprojection,
described in the previous section. First, the deprojection of both
X--ray and tSZ profiles is performed simultaneously, directly
obtaining the whole density and temperature profiles and their
errors. Second, besides the variance, it is also possible to compute
the correlation matrix for all parameters, without any extra
computational cost. Finally, it is possible to introduce in the
likelihood extra terms in order to improve the accuracy and robustness
of the technique. As we shall describe in the following, we adopt a
regularization technique, which is based on imposing a suitable
constraint to the likelihood function, to smooth out spurious
oscillations in the recovered profiles induced by the covariance in
the parameter estimate.

The definition of the likelihood is the most important part of the
whole procedure. We define a joint likelihood for the tSZ profile,
$\mathcal{L}_{tSZ}$, and for the X--ray surface brightness profile,
$\mathcal{L}_{Xray}$, also including a term associated to the
regularization constraint, $\mathcal{L}_{reg}^\lambda$. Since these
three terms are independent, the total likelihood is given by the
product of the individual ones: \be
\mathcal{L} \equiv \mathcal{L}_{tSZ} \cdot \mathcal{L}_{Xray} \cdot
\mathcal{L}_{reg}^\lambda. 
\ee 
For both the tSZ and the X--ray profiles, the take the Gaussian form
for the likelihood,
\be
ln(\mathcal{L}_{tSZ,X-ray}) = - {1 \over 2}\chi^2 = - {1 \over 2} \sum_i \left(
  {O_i - M_i \over \sigma_i} \right) ^2,
\ee 
where $O_i$ are the values of the profiles, in the $i$-th bin,
measured from the maps, while $M_i(x)$ are the model--predicted
profile values, as obtained for the set $x$ of parameters. Finally,
$\sigma_i$ is the uncertainty on the measured values $O_i$.

While the Gaussian expression is adequate for the tSZ signal, its
application for the X--ray, instead of the Poisson distribution,
requires the number of photons sampling the surface brightness map in
each radial shell to be large enough to neglect the Poisson noise. As
we shall discuss in the following, even in the outermost rings, we
always have at least 20 photons in the ``noisy'' X--ray maps.

For the regularization constraint, we adopt the Philips-Towmey
regularization method \citep[][and references
therein]{1995A&AS..113..167B}. This method has been already used also
by \cite{2006A&A...459.1007C} to deproject X--ray imaging and spectral
data. The method consists of minimizing the sum of the squares of the
$k^{th}$-order derivatives around each data-point, so as to smooth out
oscillations in the profiles. Here we choose to minimize the
second--order derivative, since we aim to eliminate fluctuations in
the profiles, but not the overall gradient. As we shall discuss in the
following, such oscillations are due either to genuine substructures
or to noise which propagates from adjacent bins in the
deprojection. The local derivative of the function $x_i$ at the $i$-th
radial interval is computed by fitting it value and the values at the
adjacent points, $x_{i-1}$ and $x_{i+1}$), with a second order
polynomial.  Let $r_i$ be the value of the equally--spaced
cluster--centric distances, at which the profiles are sampled, and
$\Delta_r$ the spacing. Then, the regularization likelihood can be
cast in the form
\ba 
ln(\mathcal{L}_{reg}^\lambda) = - {1 \over 2} \lambda'
\sum_{i=3}^{N-1} \left({ 2f_i -f_{i-1} -f_{i+1} \over 
\Delta_x^2 } \right)^2 \equiv \nonumber\\
\equiv - \lambda \sum_{i=3}^{N-1} \left({ 2f_i
-f_{i-1} -f_{i+1}}\right)^2
\label{eq:like_reg}
\ea
The quantity between parenthesis in the line formula is the exact
value of the local second--order derivative around $r_i$. All the
constant factors are included in the coefficient $\lambda$, which is
called the {\it regularity parameter}. The choice of its value is
determined by the compromise one wants to achieve between the fidelity
to the data (low $\lambda$) and the regularity of the solution (high
$\lambda$).  A small $\lambda$ value will give an inefficient
regularization, while a too high $\lambda$ will force the profile to a
straight line, especially if the signal-to-noise ratio, S/N, is
low. We apply the regularization constraint only to the temperature
profile, which is that generally showing large oscillations, while the
density profile has always a rather smooth shape. The sum in
eq.(\ref{eq:like_reg}) starts from $i=3$ since we prefer to exclude
the innermost point from the regularization procedure.

With this approach, the values of the 3D gas density and temperature
profiles are computed at $N=15$ radii each. Therefore, the total
number of parameters to be determined with the maximum likelihood approach
is 30. In order to optimize the sampling of such a large parameter
space, we adopt a Markov Chain Monte Carlo (MCMC) fitting technique
\citep[][]{1993Neal,gilks1996,mackay1996}. 

What the MCMC computes is the (marginalized) distribution of each
parameter of a set, $x_i$ (i.e. the values of density and temperature
into each bin), for which the global (posterior) probability $P(x)$,
which is proportional to the likelihood function, is known at any
point in the parameter space. In the case of an high number of
parameters or of a particularly complex $P(x)$, this is quite
difficult to be done analytically, or simply computationally very
expensive. Instead, the MCMC performs the exploration of the parameter
space with a limited
computational cost, thanks to an iterative Monte Carlo approach, by
sampling the $P(x)$ distribution. At
each iteration, new values of the parameters are drawn from a symmetric
proposal distribution, that in our case is a Gaussian,
\be
q(x_i,\hat x_i) \propto e^{-{(x_i- \hat x_i )}^2 / 2\alpha_i^2}\,.
\label{eq:distr}
\ee 
Here $x_i$ and $\hat x_i$ are the entries of two vectors, having
30 components each, which represent the updated and the old values of
the fitting parameters, respectively.  The parameter $\alpha_i$
determines the possible range for $x_i$ given $\hat x$ .

After the likelihood function is computed for a new set of parameters
$x$, these new values are accepted or rejected with a probability (A)
given by the so--called Metropolis criterion \citep{metropolis1953}:
\be\label{eq:metropolis}
A(x,\hat x)=min\left\{ 1,{P(x) \over P(\hat x)}\right\} ,
\label{eq:accept}
\ee 
where $P(x)$ is the distribution sampled by the MCMC
\footnote{\cite{hastings1970} has generalized this treatment to
non--symmetric proposal distributions, by adding a factor in equation
\ref{eq:metropolis} which takes into account the proposal distribution
$q(x,\hat x)$: 
\be
A(x,\hat x)=min\left\{ 1,{P(x) \over P(\hat x)}{q(x,\hat x) \over q(\hat
x,x)}\right\} ,
\ee
This is called the Metropolis--Hastings criterion.}. 

The width of the proposal distribution appearing in eq.(\ref{eq:distr}),
$\alpha_i$, determines the behavior of the chain: a small value of
$\alpha_i$ increases the acceptance rate since the new proposed value
is close to the old one, while a high value provides a faster
exploration of the parameter space. Our criterion to choose the values
of $\alpha_i$ is that the resulting acceptance rate, given by
eq.(\ref{eq:accept}), is around 10 per cent.

Each parameter is allowed to vary within a finite interval, in order
to avoid that the MCMC finds secondary maxima in unphysical regions of
the parameter space. As for gas density, we allow it to vary within a
large range, $0.1 < n_e < 10^{-6}$ cm$^{-3}$. Since density is mostly
constrained by the X--ray signal, which is proportional to $n_e^2$, it
is always fairly well constrained and the above large interval of
variation does not create convergence problems in any of our objects.
The upper and lower limits allowed for the temperature are 25 keV
(never reached along the chain) and 0.5 keV.  Even though none of our
clusters reach such low temperatures within the virial radius, the
exploration of the parameter space during the Markov Chain run could
reach such a low temperature regime. When this happens, the rapid drop
of the cooling function $\Lambda(T)$ below 0.5 keV generates a maximum
in the likelihood probability distribution, with unphysically low
temperature and very high density.

The iterative procedure described above is repeated until a suitable
number of new sets of parameters are accepted in the chain (typically
$\magcir 5\times 10^4$). In this condition, the frequency of the
occurrence in the chain of the $i$-th parameter $x_i$ approaches its
true probability distribution, $P(x_i)$. Note that each parameter
distribution is already marginalized over the distributions of all the
other parameters.

We perform the statistical analysis of the chain by using the code
{\it getdist} of the COSMOMC package \citep[]{2002PhRvD..66j3511L}.
In addition to a complete statistical analysis of the chain, the code
performs a series of convergence tests: the Gelman \& Rubin R
statistics \citep{gelman1992}, the Raftery \& Lewis test
\citep{raftery1992} and a split--test (which essentially consists in
splitting the chain into 2, 3 or 4 parts and comparing the difference
in the parameter quantiles). We check the convergence of our result
against all these three tests.

\begin{figure*}
\centerline{
\psfig{file=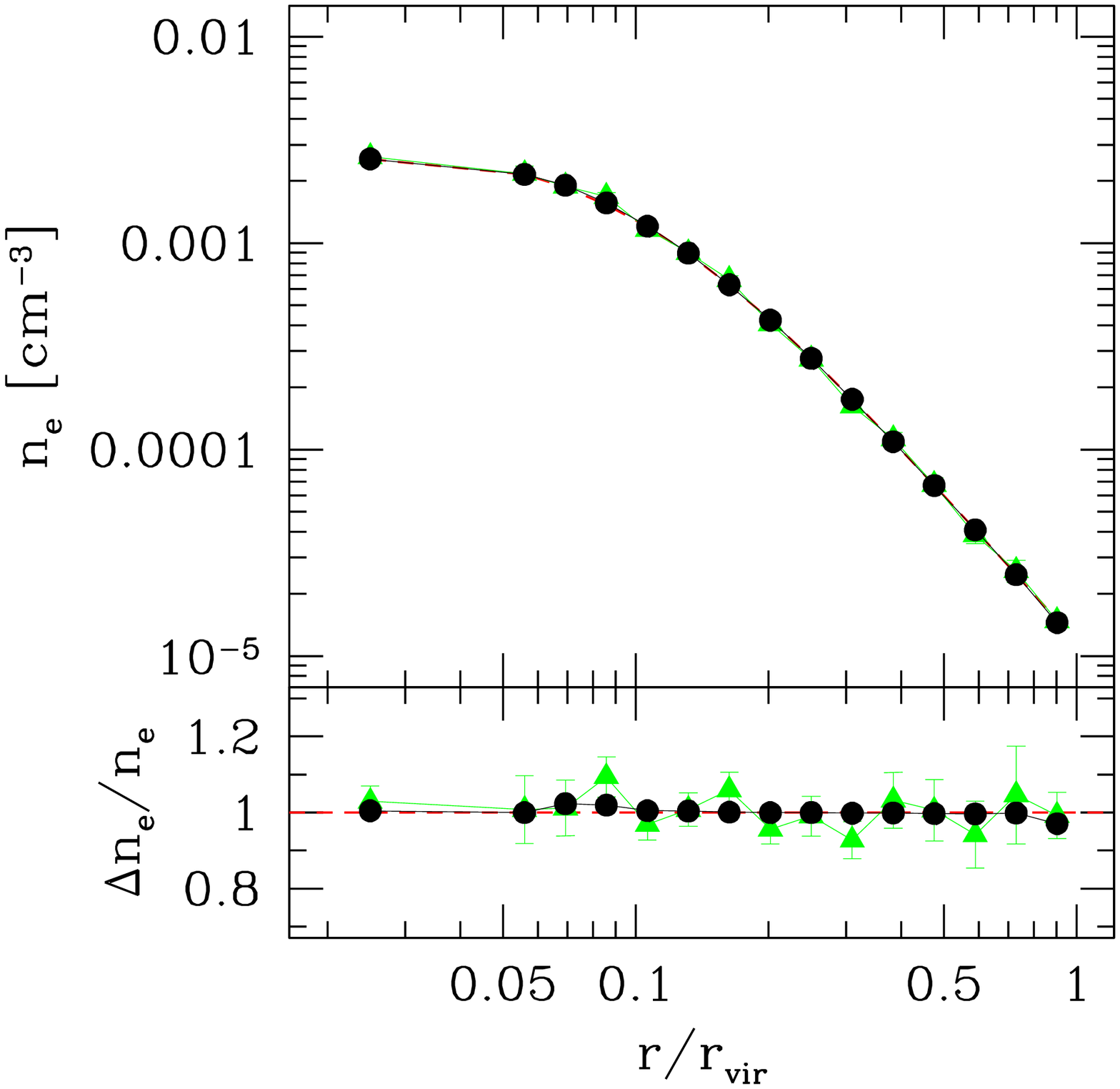,width=8.cm}
\psfig{file=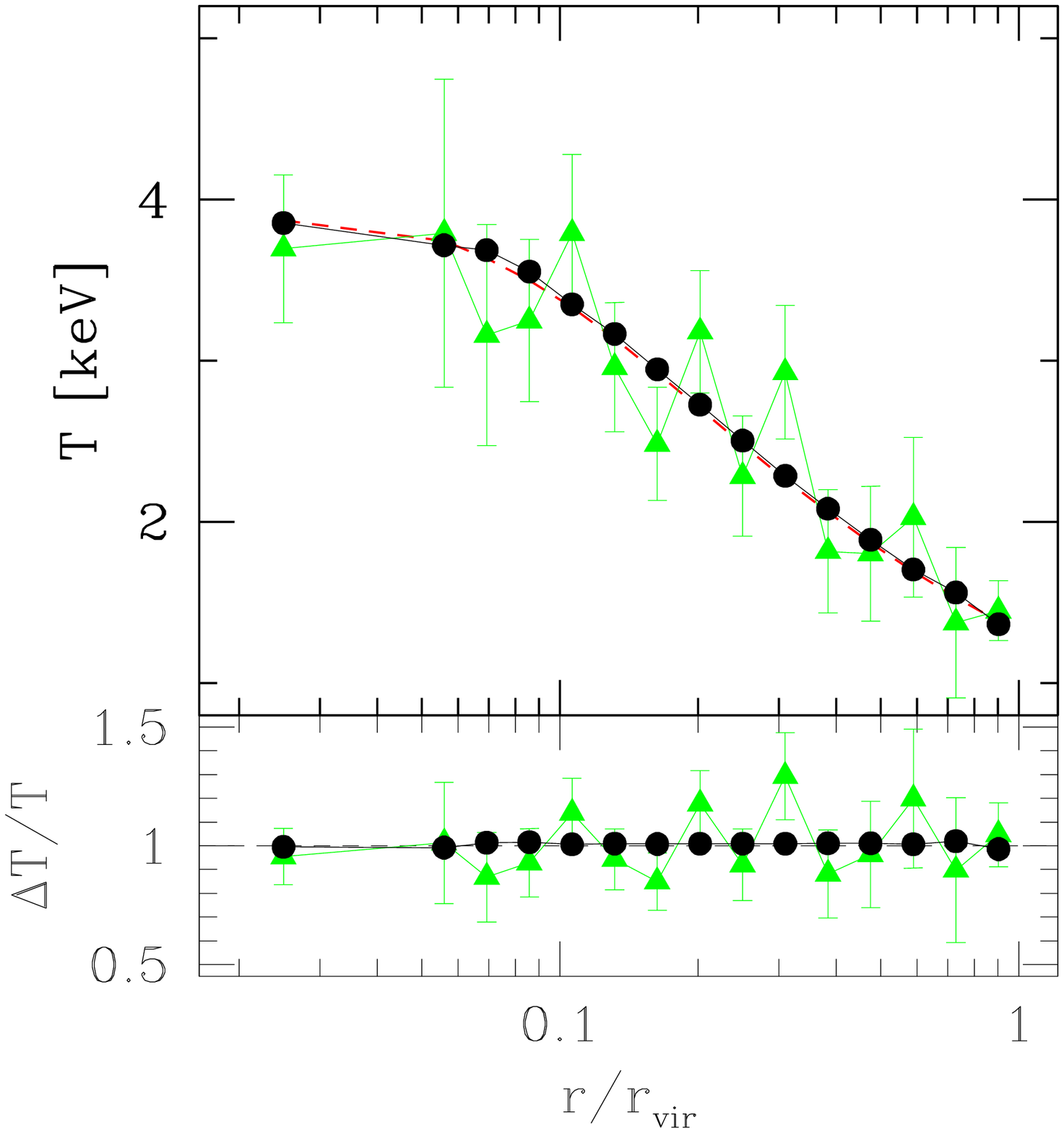,width=8.cm}
}
\caption{Results of the geometrical deprojection of the analytical
model: density profile (left panel), temperature profile (right
panel). The dashed line represents the true profile.  The circles
(triangles) connected by a solid line represent the reconstructed profiles
without (with) the inclusion of the noise. Errorbars represent
1$\sigma$ deviations over 1000 Monte Carlo resamplings over the data
(see text for details, Section \ref{sec:dpj})}
\label{fi:beta_ide_dpj}
\label{fi:beta_noi_dpj}
\end{figure*}

\section{Application to an idealized cluster model}

In order to investigate the presence of possible systematics in the
geometrical deprojection technique, we carry out a test on an ideal
cluster model. We construct this model cluster by assuming the
$\beta$--model for the gas density profile, with an effective
polytropic equation of state to define the temperature profile: 
\ba
n_e & =& n_{e0}\left( 1+ {r^2\over r_c^2}\right)^{-3\beta/2} \nonumber
\\ T & = & T_0 \left({n_e\over n_{e0}}\right)^{\gamma - 1} = T_0
\left( 1+ {r^2\over r_c^2}\right)^{-3\beta(\gamma-1)/2} 
\ea 
The values of the model parameters are fixed as follows: $\beta =
0.8$, $\gamma =1.2$ for the effective polytropic index, $n_{e0} =
3\cdot 10^{-3}$ cm$^{-3}$ for central electron number density, $T_0 =
4$ keV for the central temperature, $r_c = 200$ kpc for the core
radius. The ``virial'' radius, which represents here the largest
cluster-centric distance out to which the profiles are followed, is
fixed at $r_{vir} = 2$ Mpc. We assume the cluster to be placed at
redshift $z=0.1$.

We create tSZ and X--ray images of this model in X--ray and tSZ. The
maps are composed by a grid 512x512 pixels. The physical dimension of
the pixel is $\sim 16$ kpc, which corresponds to an angular scale of
$\sim 4$ arcsec at this redshift. In realizing the map of X--ray
surface brightness, we adopt for the cooling function the pure
bremsstrahlung expression $\Lambda (T) = \Lambda_0 (T/T_0)^{0.5}$
where $\Lambda_0 = 5\cdot 10^{-23}$ erg/s cm${^3}$. We prefer this
simple formula since at this stage our main interest is to investigate
the systematics of the deprojection itself, rather than the
uncertainties introduced by the dependence of the X--ray emissivity on
the temperature (which we expect anyway to be quite small).

We adopt the same binning strategy for both the ideal cluster and for
the simulated objects, that we shall describe in the next Section.
The first bin is taken from $r=0$ to $r=0.05r_{vir}$ which always
corresponds to \magcir 100 kpc in our set of simulated clusters. Then,
we compute the profile in 10 (15) bins out to $R_{500}$ ($R_{vir}$)
which are equally spaced in logarithm. This choice represents a good
compromise between the needs of accurately resolving the profile and
of having an adequate signal-to-noise (S/N \magcir 5) in each bin.  We
point out that a proper binning criterion is important in order to get
an unbiased reconstruction of the profiles.  One should keep in mind
that the spherical shells are assumed to have homogeneous gas density
and temperature structures, thus neglecting any internal radial
gradient. On the other hand, the portion of each shell, which is
projected on the corresponding ring in the image, is located at a
larger radial distance from the center than the portion of the same
shell which is projected into the inner rings. Therefore, if the bin
width is comparable or larger larger than the scale length of the
internal radial gradient of the shell, the emissivity contributed to
the correspondent ring is lower than expected from a homogeneous gas
density, while it is larger for all the inner rings. As a consequence,
the emissivity of the shell is underestimated while that of all the
inner shells is overestimated in order to correctly reproduce the
cluster image.

In order to check for the presence of such systematics, we first apply
the deprojection technique in the case of an ideal observation, free
of any noise. The reconstructed density and temperature profiles are
shown in Figure \ref{fi:beta_ide_dpj}.  The reconstruction in this
extremely idealized case is excellent, with very small or no
deviations in all bins. Larger deviations are in the outermost bins
and are related to the subtraction of the contribution of the
fore--background contaminations. This contamination is due to the fact
that the $\beta$--model used to produce our maps ideally extends out
to infinity. Nevertheless, all deviations are smaller than a few
percent and are negligible with respect to any observational
noise. The results obtained in this test case show that taking equally
log-spaced bins is in fact a good choice. 

\subsection{Geometrical deprojection of the noisy maps} \label{sec:dpj}
\label{par:noise}
The case of noiseless observations discussed in the previous section
is highly idealized. The impact of including a realistic noise is
instead very important and cannot be neglected. The recipe to add
noise to the maps, that we describe here, will also be used in the
study of the simulated clusters, discussed in Section 4. 

As already mentioned in the Introduction, recovering detailed
temperature profiles from the combination of X--ray imaging and tSZ
data requires both of them to have an adequate spatial
resolution. While this is certainly the case for the present
generation of X--ray satellites, the combination of good sensitivity
and spatial resolution for tSZ observations should await the next
generation of sub-millimetric telescopes. For the purpose of our
analysis, we model the noise in the tSZ maps by using as a reference
the performances expected for the planned Cornell--Caltech Atacama
Telescope (CCAT), which is expected to start operating at the
beginning of the next decade\citep{2006SPIE.6267E..75S}. The telescope
will be a single--dish with 25 m diameter. The required
field--of--view is of about $10\times 10$ arcmin$^2$, with the goal of
covering a four times larger area, so as to cover one entire rich
cluster down to a relatively low redshift. The best band for tSZ
observations will be centered on 150 GHz. At this frequency CCAT is
expected to have Gaussian beam of 0.44 arcmin FWHM. The first step of
noise setup is to convolve the maps with this beam. Then we add a
Gaussian noise of 3$\mu$K/beam. This level of noise should be reached
with about 6 hours of exposure with CCAT.  In the present study, we
neglect in the tSZ maps any contamination, in particular we do not
consider the presence of unresolved radio point sources. A detailed
analysis of the contaminations in the tSZ signal has been provided by
\cite{2004ApJ...612...96K} and by \cite{2004astro.ph..2571A}.

As for X--ray observations, the Chandra satellite is currently
providing imaging of superb quality, with a sub--arcsec resolution on
axis. A proper simulation of X--ray observations should require
generating spectra for each pixel, to be convolved with the response
function of a given instrument. However, in order to apply our
reconstruction method we only need to generate X-ray surface
brightness maps with a given number of events (photons), regardless of
their energy. For this purpose we simulate the X--ray photon counts by
using a Monte Carlo sampling of the surface brightness map.  We fix to
$N=10^4$ the total number of photons within the virial radius of the
cluster, which is quite typical for medium--deep observations of
relatively nearby clusters. Each photon event is generated in a
particular pixel $i$, with probability
\be
P_i={s_i \over \sum_j s_j},
\ee
where $s_i$ is the surface brightness of the pixel and the sum is
extended to all the pixels of the map. The number of expected counts
into each pixel will be given by a Poisson probability distribution
with mean $n_i=NP_i$. The conversion between counts and surface
brightness is then given by $\Sigma=\sum_i s_i / N$, so that the total flux
in the map is conserved. 

Clearly, this method of introducing noise in the X--ray maps  only
takes into account the statistical errors associated to finite
exposures. However, it neglects the effects of any systematics (e.g.,
contribution of the instrumental and cosmic background, etc.) which
should be included in a more realistic observational setup. A
comprehensive description of the instrumental effects on the recovery
of X--ray observables, calibrated on hydrodynamical simulations, has
been provided by \cite{2004MNRAS.351..505G} \citep[see
also][]{2006MNRAS.369.2013R}. Probably the most serious limitation in
our approach is that we assume the absence of any background or,
equivalently, that the background can be characterized and removed
with arbitrary accuracy.

\begin{figure}
\centerline{
\psfig{file=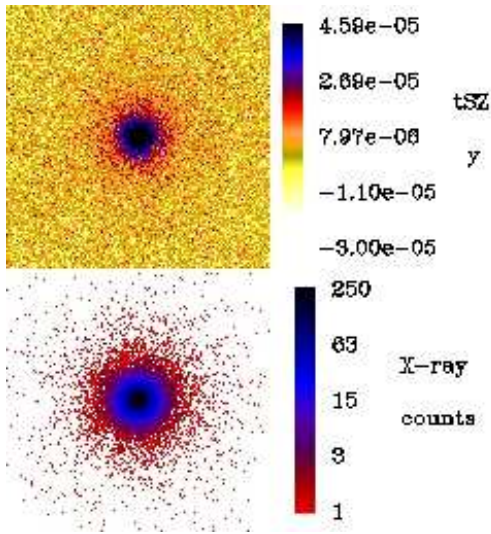,height=7cm}
}
\caption{tSZ (upper panel) and X--ray (lower panel) maps of the ideal
cluster.  The side of the map corresponds to $2R_{vir}=4$ Mpc. In the
bottom left corner of the tSZ map is the beam. Note that the scale is
linear in the tSZ map and logarithmic in the X--ray one.}
\label{fig:ide_maps}
\end{figure}

In Figure \ref{fig:ide_maps} we show are the tSZ and X--ray images of
the idealized cluster, once noise is added as described above.  In
Figure \ref{fi:beta_noi_dpj} we show the results of the deprojection
of the noisy maps of the idealized cluster, for both the density and
the temperature profiles. In order to estimate the errors in the
deprojected profiles, we perform a MonteCarlo resampling of the
projected X--ray and tSZ profiles: the value of the profile within each
radial ring is randomly scattered according to a Gaussian
distribution, whose width is given by the error associated to the
noise introduced in the map.  The 1$\sigma$ errors in the deprojected
profiles is then obtained as the scatter within a set of 1000
deprojections of the MonteCarlo--resampled tSZ and X--ray profiles.

The density is the best determined quantity, with uncertainty lower
than 10 per cent. This is quite expected, owing to the sensitive
dependence on gas density of both the X--ray signal ($\propto n_e^2$)
and of the tSZ one ($\propto n_e$). The temperature has instead higher
errors, of about 20--30 per cent. This is due to the fact that the
X--ray signal has a weaker dependence on the temperature (only
contained in the cooling function $\Lambda(T)$). For this reason, the
determination of the temperature profile, independent of any X--ray
spectroscopic analysis, is strictly related to the possibility of
having high--quality tSZ data.

The introduction of noise generates fluctuations in the tSZ and X--ray
profiles which translate into variations of the recovered density and
temperature. Looking at the bottom panels of
Fig. \ref{fi:beta_noi_dpj}, positive fluctuations in the density
correspond to negative fluctuations in the temperature (and
viceversa). Furthermore, any fluctuation in a given direction in one
radial bin generally corresponds to a fluctuation in the opposite
direction in an adjacent bin, within the same profile.  This pattern
in the fluctuations witnesses the presence of a significant covariance
among nearby bins in the same profile and between the values of
density and temperature recovered within the same radial bin. As for
the covariance between neighbor bins, it is due to the onion--skin
structure assumed in the deprojection: every time that a quantity is
over(under)estimated in a radial bin, the deprojection forces the same
quantity to be under(over)estimated in the adjacent inner bin, so as
to generate the correct projected profile.  As for the covariance
between different profiles, it is mostly induced by the tSZ signal,
which has the same dependence on both $n_e$ and $T$.  Although such
oscillations are present for both density and temperature, they are
smaller for the former, due to its faster decrease with radius.

\subsection{Maximum likelihood deprojection of the noisy maps}

\begin{figure*}
\centerline{
\psfig{file=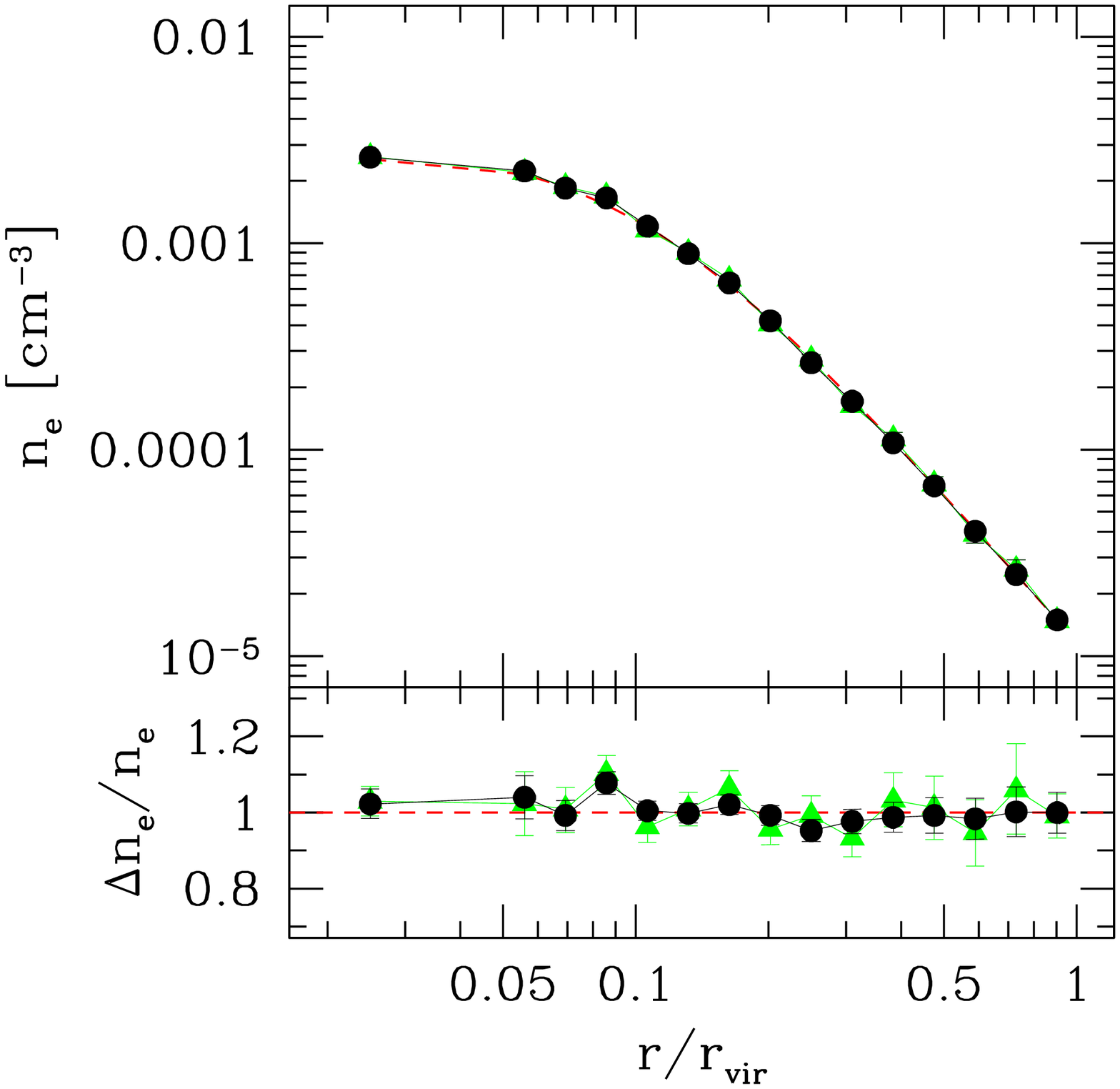,width=8.cm}
\psfig{file=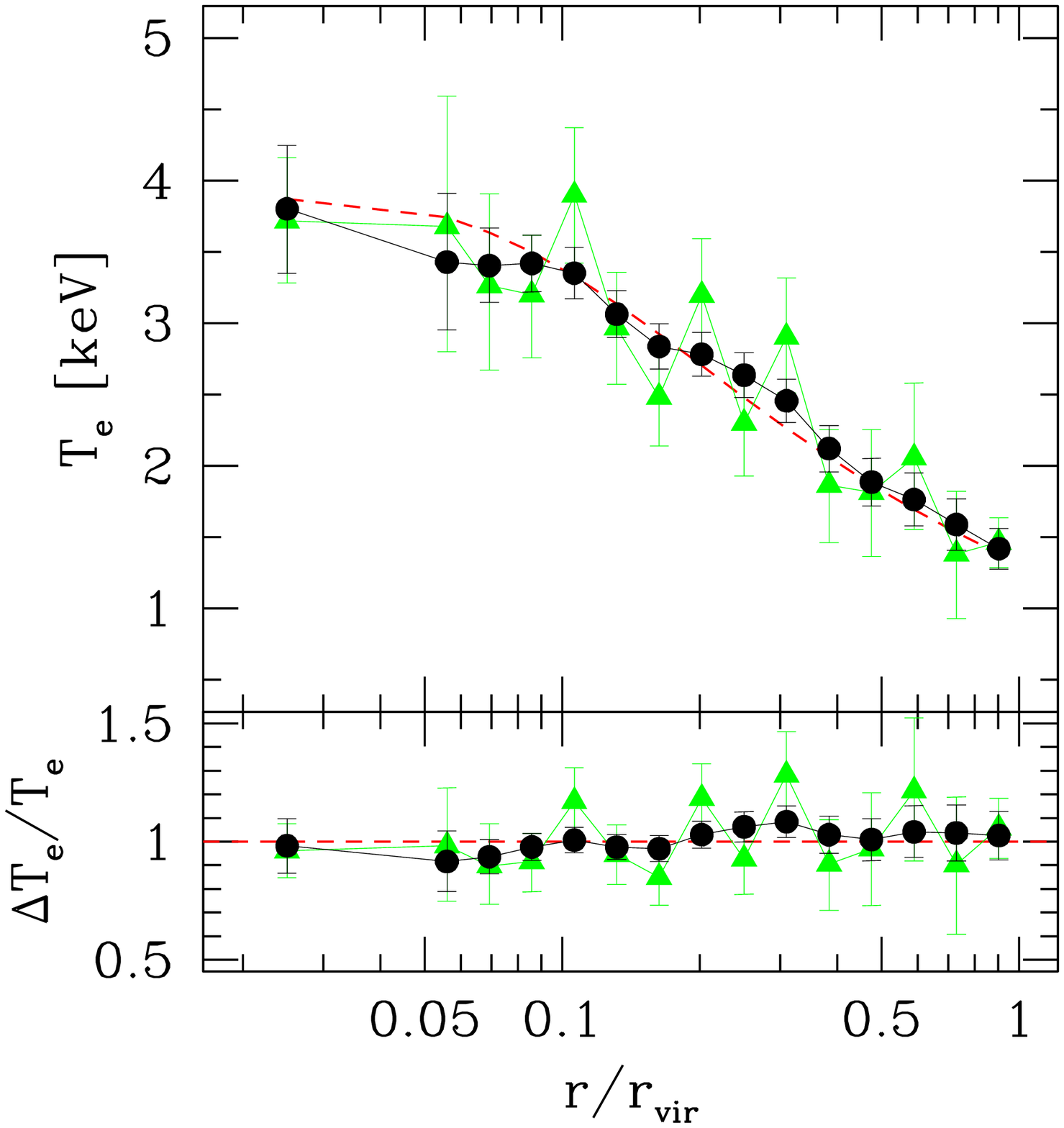,width=8.cm}
}
\caption{Results of the maximum--likelihood deprojection on the
  density profile (left panel) and temperature profile (right panel)
  for the idealized cluster. In the upper part of each panel we
  show the correct profile (dashed curve) and the reconstructed
  profile with and without the regularization constraint (circles and
  triangles, respectively). In the lower part we show the fractional
  deviation of the recovered profiles from the true one. }
\label{fi:beta_noi_jof}

\end{figure*}

We verified that using the maximum--likelihood technique, as described
in Sect. 2.2, generally produces very similar results to those of the
geometrical deprojection, at least when the regularization term,
$\mathcal{L}_{reg}^\lambda$ is not included in the analysis. The
results of this deprojection method on the polytropic $\beta$-model
are shown in Figure \ref{fi:beta_noi_jof}, where we also show the
effect of introducing the regularization term.  The effect of the
regularization constraint is evident: most of the fluctuations, which
are due to the degeneracies between fitting parameters, disappear and
the deprojected profiles become much more regular, and with smaller
errorbars in the profiles, while the accuracy of the reconstruction
remains essentially unbiased. 

\begin{figure*}
\centerline{
\psfig{file=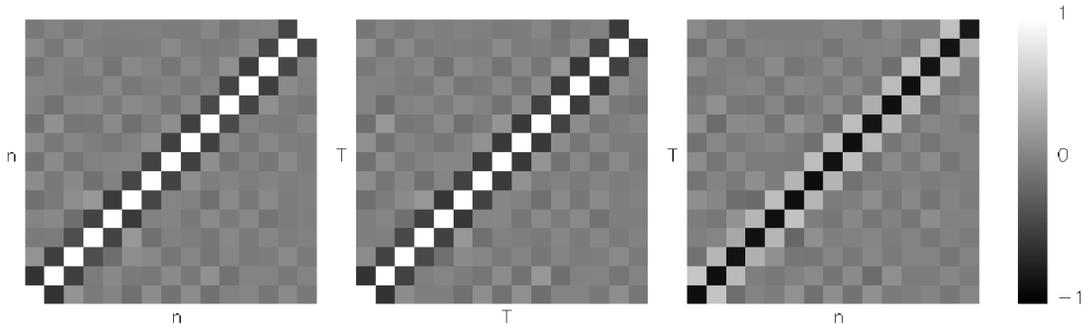,width=15.cm}
} 
\caption{The correlation matrix of density and temperature without
the regularization constraint: density-density (left panel),
temperature-temperature (central panel) and density-temperature (right
panel). White pixels correspond to the presence of strong positive
correlation, while black pixels are for strong anti--correlation.}
\label{fi:beta_corr_old}
\end{figure*}

\begin{figure*}
\centerline{
\psfig{file=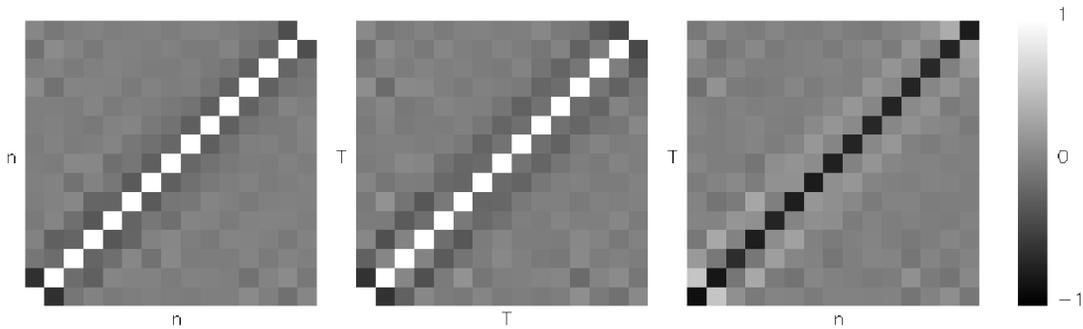,width=15.cm}
} 
\caption{The correlation matrix of density and temperature while using
regularization constraint: density-density (left panel),
temperature-temperature (central panel) and density temperature (right
panel). The color-coding of the pixels is the same as in Figure
\ref{fi:beta_corr_old}.} 
\label{fi:beta_corr}

\end{figure*}

In order to study in detail the presence of correlations among the
fitting parameters, we compute the correlation matrix, which is
defined as $C_{ij}=\sigma_{ij}/\sigma_i \sigma_j$, where $\sigma_{ij}$
is the covariance between the $x_i$ and the $x_j$ fitting parameters,
while $\sigma_i^2$ is the variance for the $x_i$ parameter. The
covariance matrix is computed along the Markov Chain. Therefore,
$C_{ij}$ is in our case a matrix with $30\times 30$ entries. In
Figures \ref{fi:beta_corr_old} and \ref{fi:beta_corr} we plot the
entries of the correlation matrix for the density--density (DD),
temperature--temperature (TT) and density--temperature (DT)
``blocks'', before and after introducing the regularization term in
the likelihood function, respectively. By definition, the variance
terms in the diagonal of the DD and TT matrices are characterized by
the maximum correlation. On the contrary, the diagonal of the DT
matrix has the maximum anticorrelation, thus demonstrating that any
positive fluctuation in the recovered profile of one quantity
corresponds to a negative fluctuation of the other quantity at the
same radius. We also note that the next-to-diagonal terms in the DD
and TT blocks have a degree on anticorrelation, thus explaining the
fluctuating profile shown in Figs. \ref{fi:beta_noi_dpj} and
\ref{fi:beta_noi_jof}. When the regularization is introduced, the
correlations between density or temperature of adjacent bins is
efficiently suppressed.

\section{Application to simulated clusters}

\begin{figure*}
\centerline{ 
\psfig{file=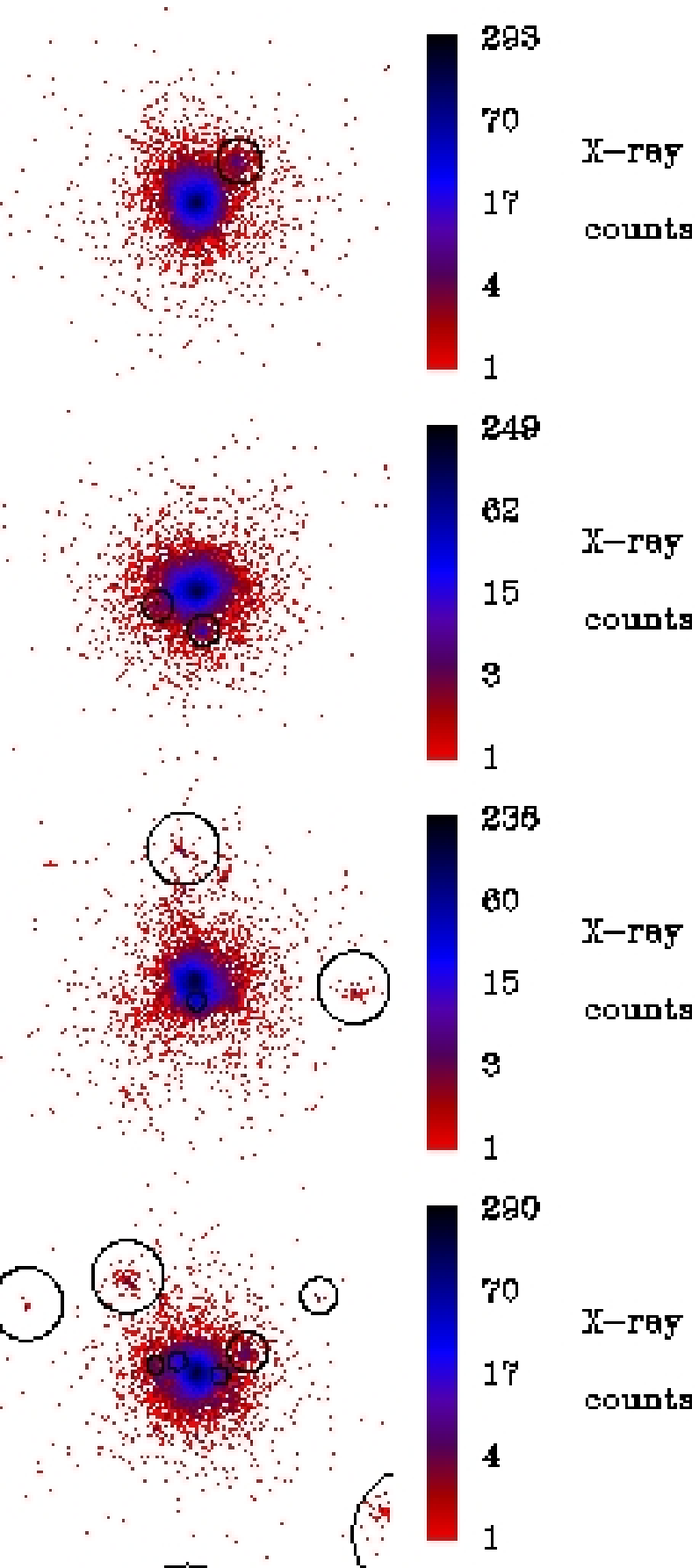,width=9.cm}
\psfig{file=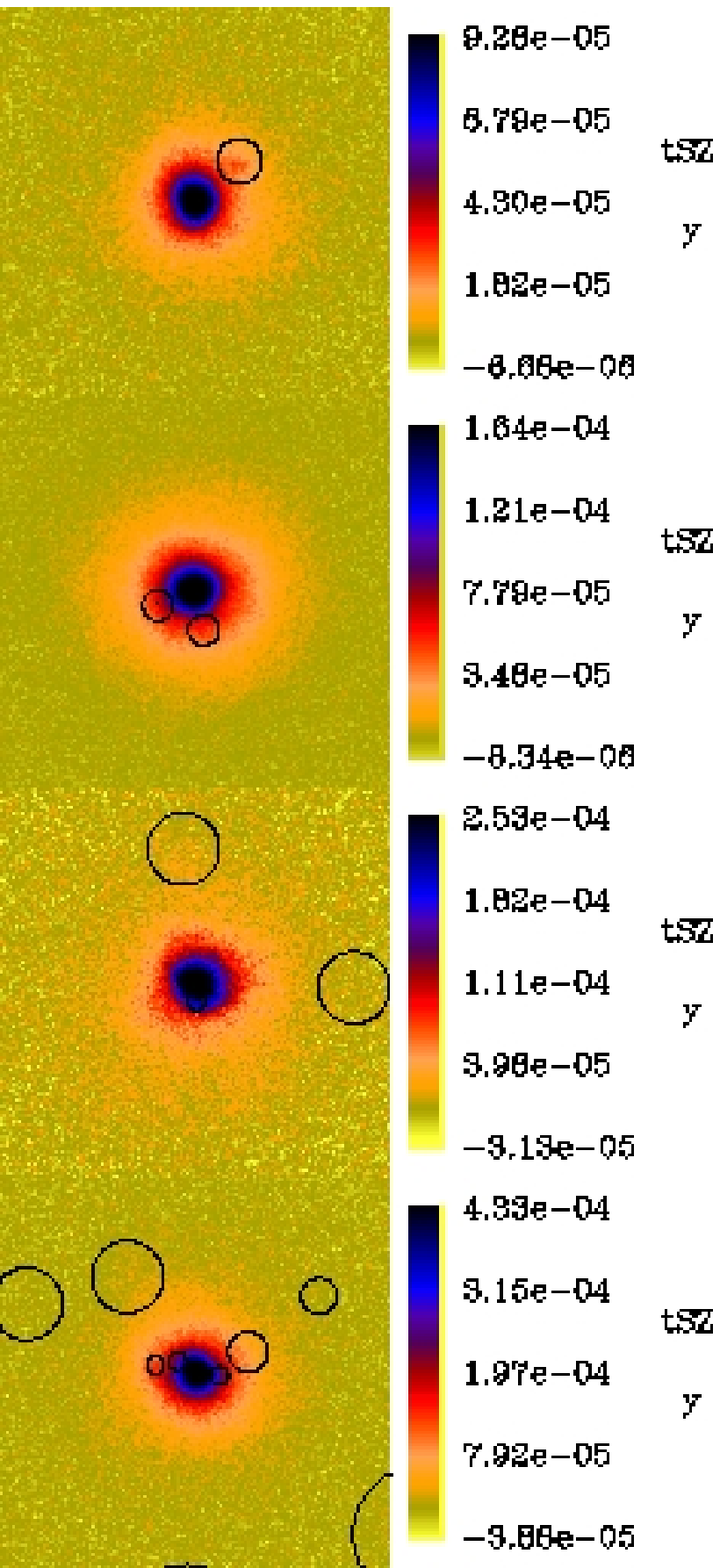,width=9.cm} 
}
\caption{Maps of the X--ray surface brightness (left panel) and of the
  Comptonization parameter (right panel) for the C1 to C4 simulated
  clusters (from top to bottom). Each map extends out to
  $R_{vir}$. Noise is added according to the recipe described in
  Section 3.1. Circles mark the regions which have been masked, due to
  the presence of detected substructures.}
\label{fi:maps_sim}
\end{figure*}

The sample of simulated galaxy clusters used in this paper has been
extracted from the large-scale cosmological hydro-N-body simulation of
a ``concordance'' $\Lambda$CDM model with $\Omega_m=0.3$ for the
matter density parameter at present time, $\Omega_\Lambda=0.7$ for the
cosmological constant term, $\Omega_{\rm b}=0.019\,h^{-2}$ for the
baryons density parameter, $h=0.7$ for the Hubble constant in units of
100 km s$^{-1}$Mpc$^{-1}$ and $\sigma_8=0.8$ for the r.m.s. density
perturbation within a top--hat sphere having comoving radius of $8\hm$
(see \citealt{2004MNRAS.348.1078B}, for further details).  The run,
performed with the Tree+SPH code {\small GADGET-2}
\citep{SP01.1,2005astro.ph..5010S}, follows the evolution of $480^3$
dark matter particles and an equal number of gas particles in a
periodic cube of size $192 h^{-1}$ Mpc. The mass of the gas particles
is $m_{\rm gas}=6.9 \times 10^8 h^{-1} M_\odot$, and the
Plummer-equivalent force softening is $7.5 h^{-1}$ kpc at $z=0$.  The
simulation includes the treatment of radiative cooling, a uniform
time--dependent UV background, a sub--resolution model for star
formation and energy feedback from galactic winds
\citep{2003MNRAS.339..289S}. The sample of clusters analyzed here
includes 14 clusters extracted at $z=0$ and having virial mass
$M_{vir} \magcir 4\times 10^{14} M_\odot$. For these clusters we
compute the electron temperature,
\be
T_{e}\,=\,{\sum_i m_i T_i\over \sum_i m_i}\,,
\label{eq:tel}
\ee
whose definition coincides with the mass--weighted one under the
assumption of a fully ionized plasma. In the above equation, $m_i$ and
$t_i$ are the mass and the temperature of the $i$-th gas particle.

Due to the finite box size, the largest cluster found in the
cosmological simulation has $T_e\approx 5$ keV.  In order to
extend our analysis to more massive and hotter systems, which are
mostly relevant for current tSZ observations, we include four more
galaxy clusters having $M_{\rm vir}>10^{15} \msun$\footnote{Here and
in the following, the virial radius, $R_{\rm vir}$, is defined as the
radius of a sphere centered on the local minimum of the potential,
containing an average density, $\rho_{\rm vir}$, equal to that
predicted by the spherical collapse model. For the cosmology assumed
in our simulations it is $\rho_{\rm vir}\simeq 100 \rho_{\rm c}$,
being $\rho_{\rm c}$ the cosmic critical density. Accordingly, the
virial mass, $M_{\rm vir}$, is defined as the total mass contained
within this sphere.}  and belonging to a different set of hydro-N-body
simulations \citep{2006MNRAS.tmp..270B}. Since these objects have been
obtained by re-simulating, at high resolution, Lagrangian regions of a
pre-existing cosmological simulation \citep{YO01.1}, they have a
better mass resolution, with $m_{\rm gas}= 1.69 \times 10^{8}
h^{-1}M_\odot$, and a correspondingly smaller softening of $5
h^{-1}$kpc at $z=0$. These simulations have been performed with the
same choice of the parameters defining star--formation and
feedback. The cosmological parameters also are the same, except for a
higher power spectrum normalization, $\sigma_8=0.9$.

\begin{table}
\centerline{
\begin{tabular}{lccc}
Cluster & $T_{e}$ & $M_{vir}$ & $R_{vir}$ \\
        & keV & $10^{14}M_{\odot}$ & Mpc \\
C1 & 2.5 & 4.0 & 2.1 \\
C2 & 4.3 & 10.1 & 2.6 \\
C3 &5.5  & 26.6 & 3.1 \\
C4 & 7.0 & 30.5 & 3.3\\
\end{tabular}
}
\caption{Characteristics of the simulated clusters, for which the
  detail of the analysis are presented. Col. 1: electron
  (mass--weighted) temperature; Col. 3: virial
  mass; Col. 4: virial radius.}
\label{tab:clusters}
\end{table}

In the following, we will show detailed results for a subset of 4
clusters. The basic characteristics of these four selected clusters
are reported in Table \ref{tab:clusters}. The first three of them are
extracted from the cosmological hydrodynamical simulation, while C4 is
one of the massive clusters simulated at higher resolution.  C2, C3
and C4 are typical examples of clusters at low, intermediate and high
temperature, while C1 is an interesting case to understand the effect
of fore--background contaminations. We show in Figure
\ref{fi:maps_sim} the X--ray surface brightness and Compton--$y$ maps
for these four clusters. All the maps are generated by placing the
cluster at redshift $z=0.1$, so that the maps, which extend out to
$r_{vir}$, have an angular size ranging from about 9 arcmin for C1 to
14 arcmin for C4.

\begin{figure*}
\centerline{ 
\psfig{file=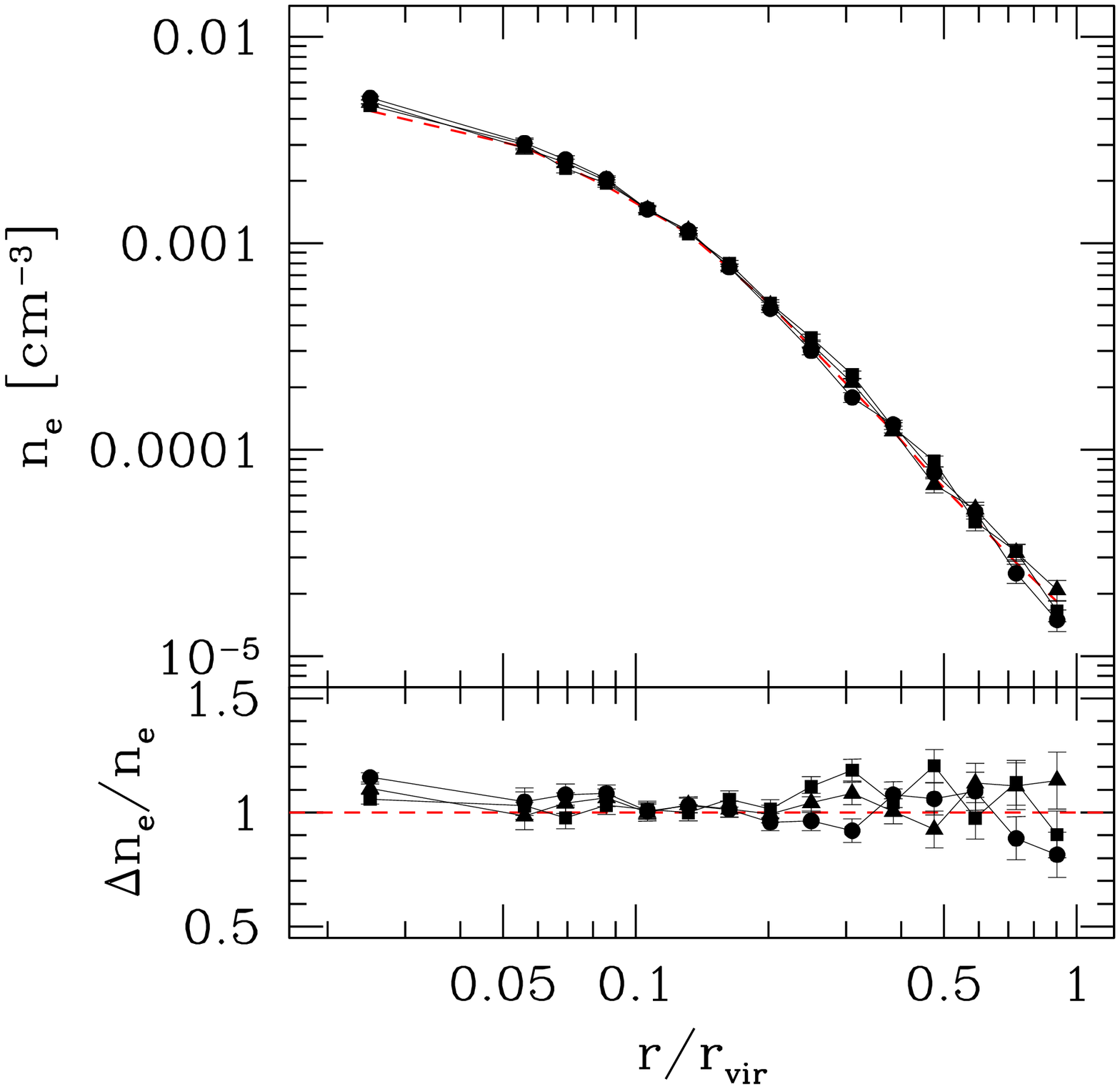,width=8.cm}
\psfig{file=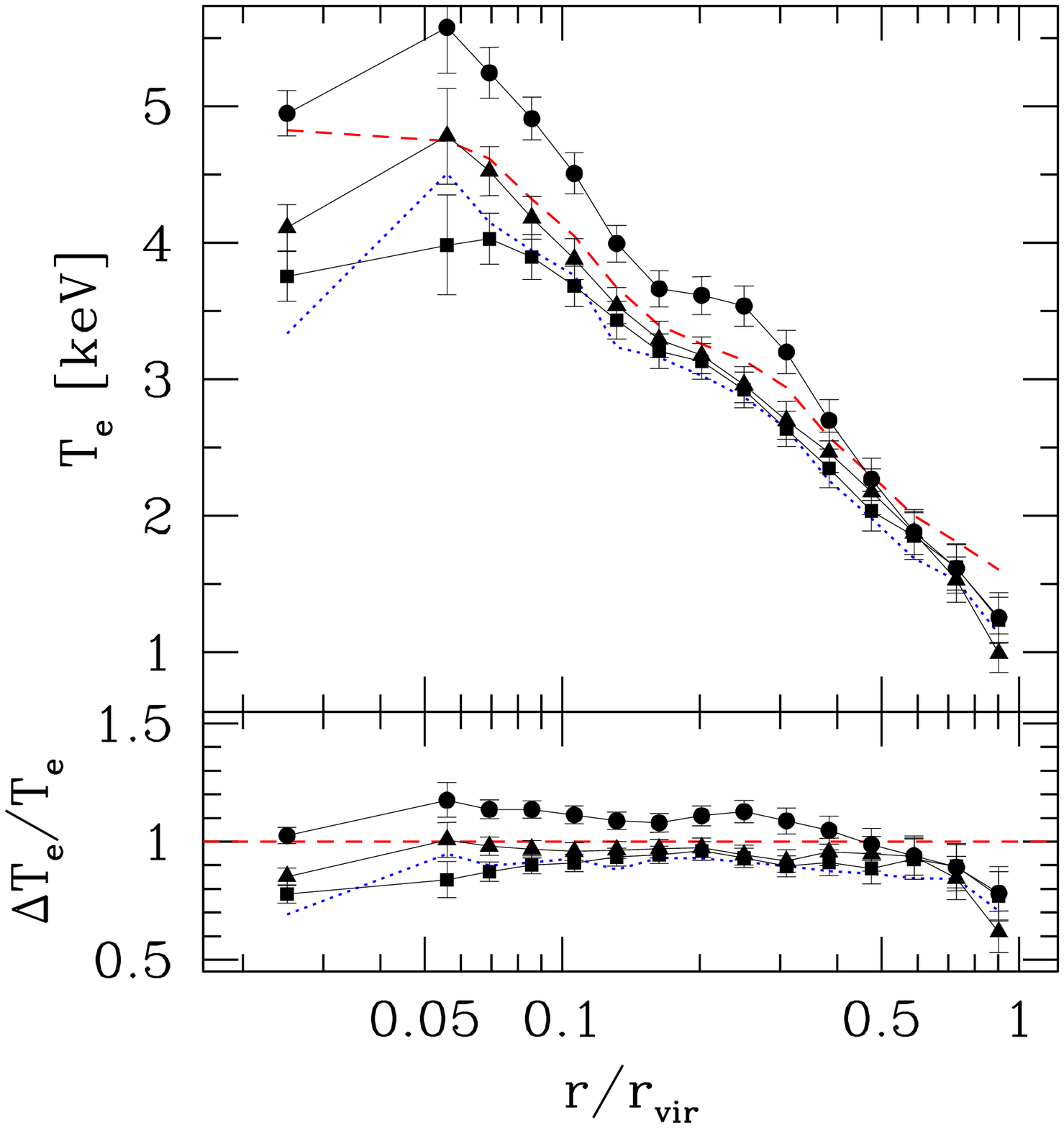,width=8.cm}
}
\caption{Application of the regularized maximum likelihood
  deprojection on the cluster C1, out to $R_{vir}$. The tree solid
  lines connecting dots with errorbars represent the reconstructed
  profile, for three orthogonal projection directions: along the $x$
  (squares), the $y$ (triangles) and the $z$ (circles) axes. Errorbars
  corresponds the asymmetric 68 per cent confidence levels, computed
  from the distribution of values taken by the likelihood function
  along the Markov Chain. The dashed line represents the true
  3-dimensional profile. The dotted line in the right panel shows the
  profile of the spectroscopic--like temperature. In the bottom
  panels, we plot the fractional deviation of the reconstructed
  profiles from the true electron temperature.}
\label{fi:c9964jof.vir}
\end{figure*}

\begin{figure*}
\centerline{ 
\psfig{file=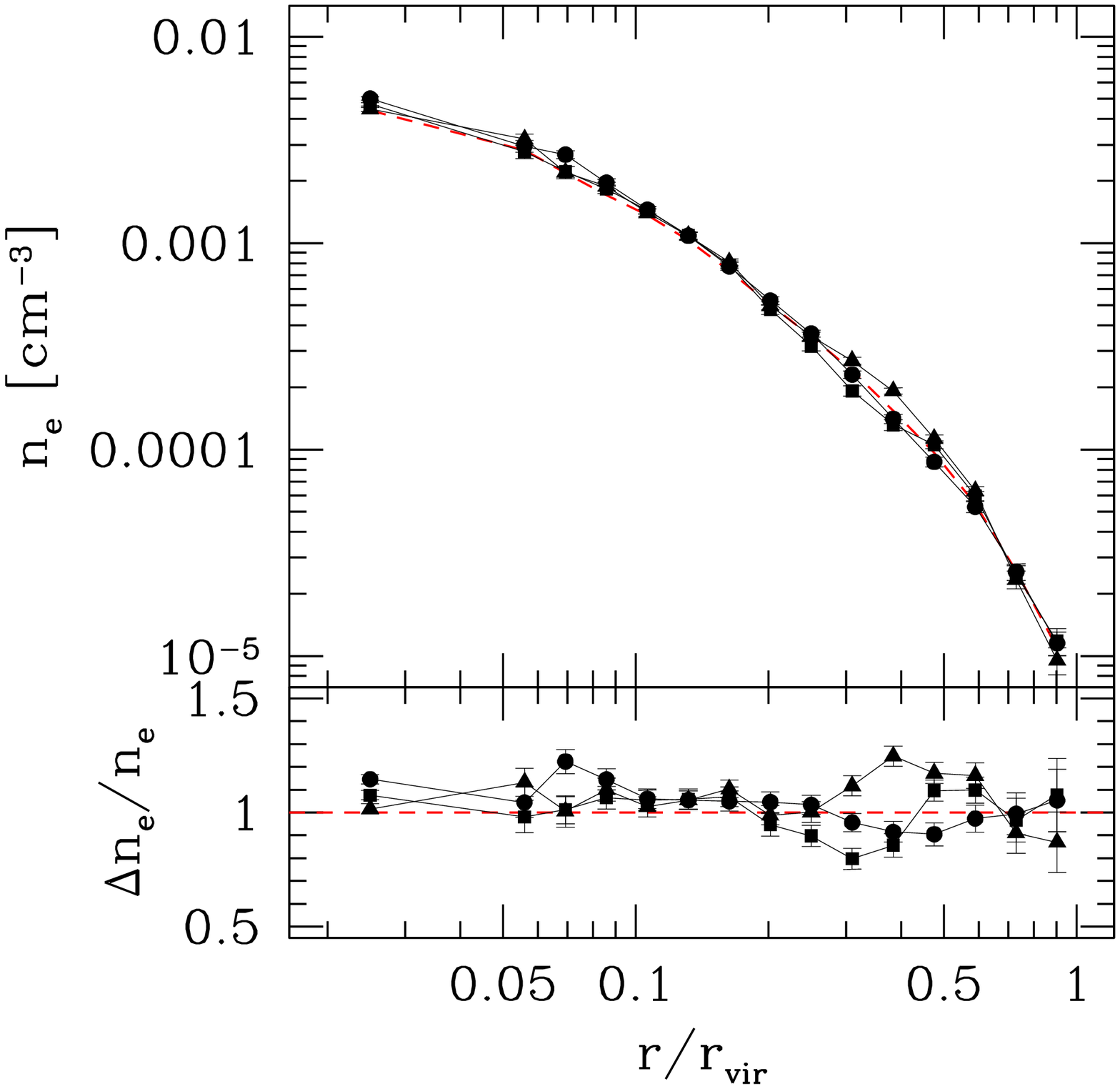,width=8.cm}
\psfig{file=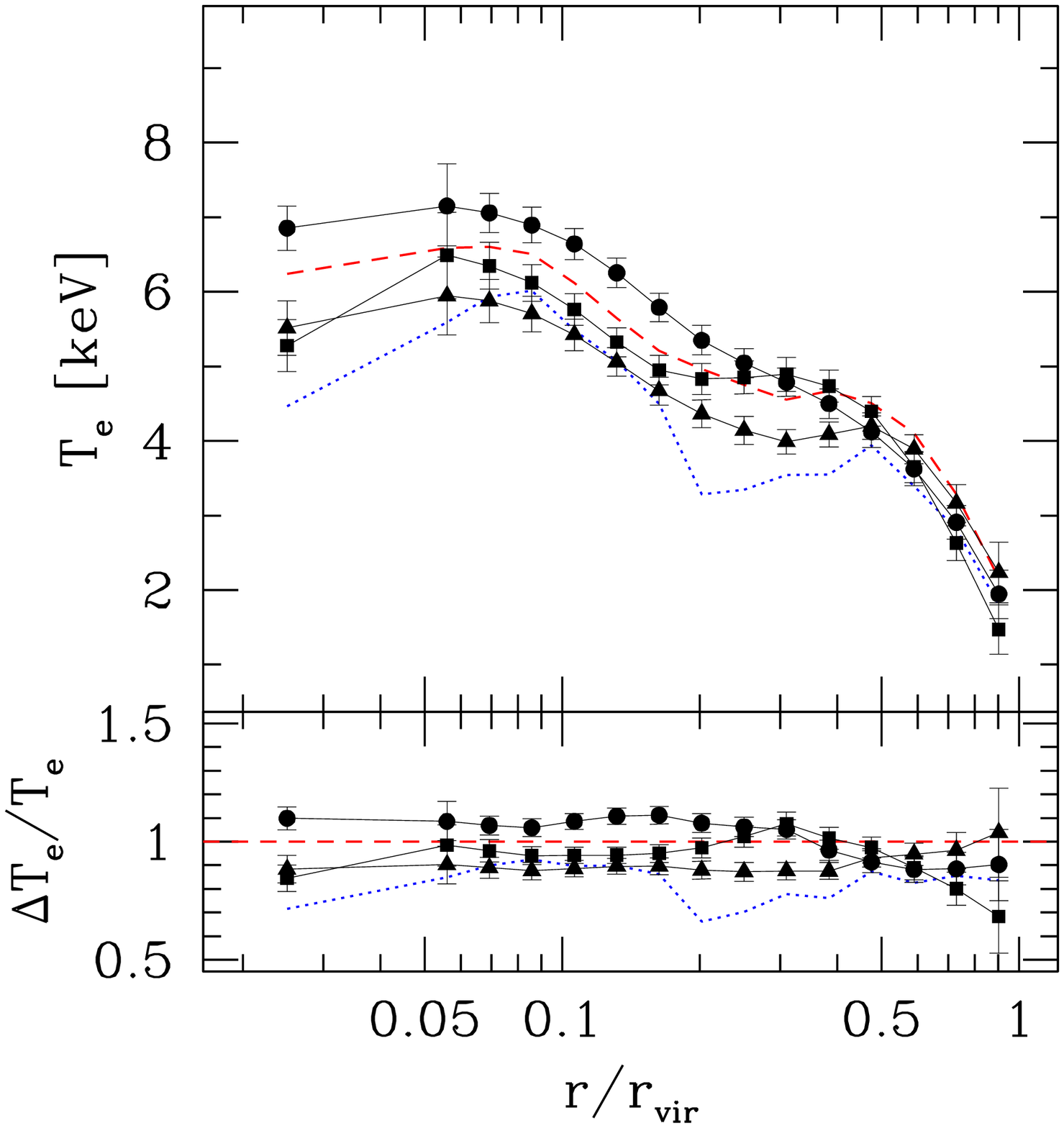,width=8.cm}
}
\caption{The same as in Figure \ref{fi:c9964jof.vir}, but for the C2
  cluster.}
\label{fi:c29931jof.vir}
\end{figure*}

\begin{figure*}
\centerline{ 
\psfig{file=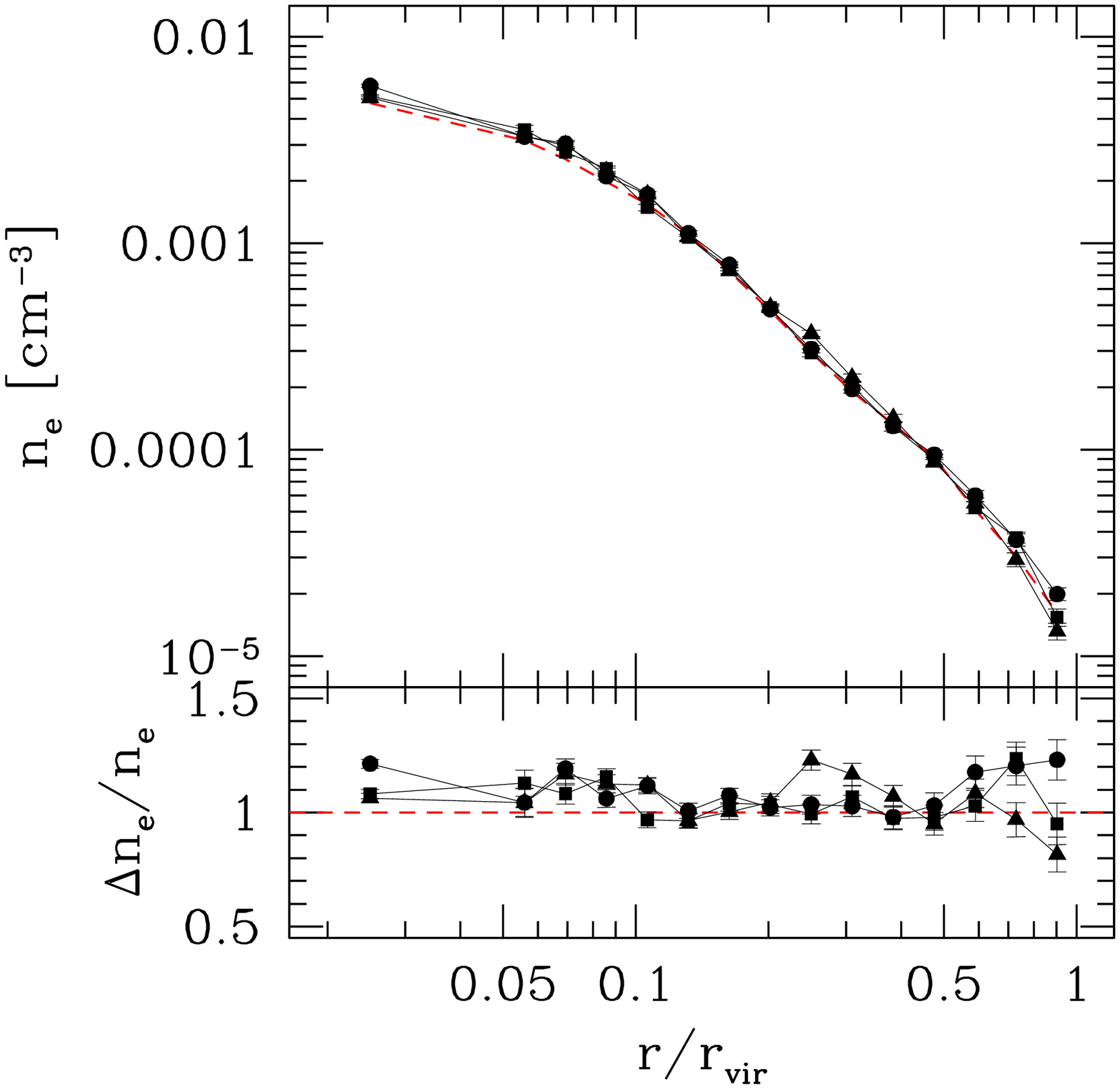,width=8.cm}
\psfig{file=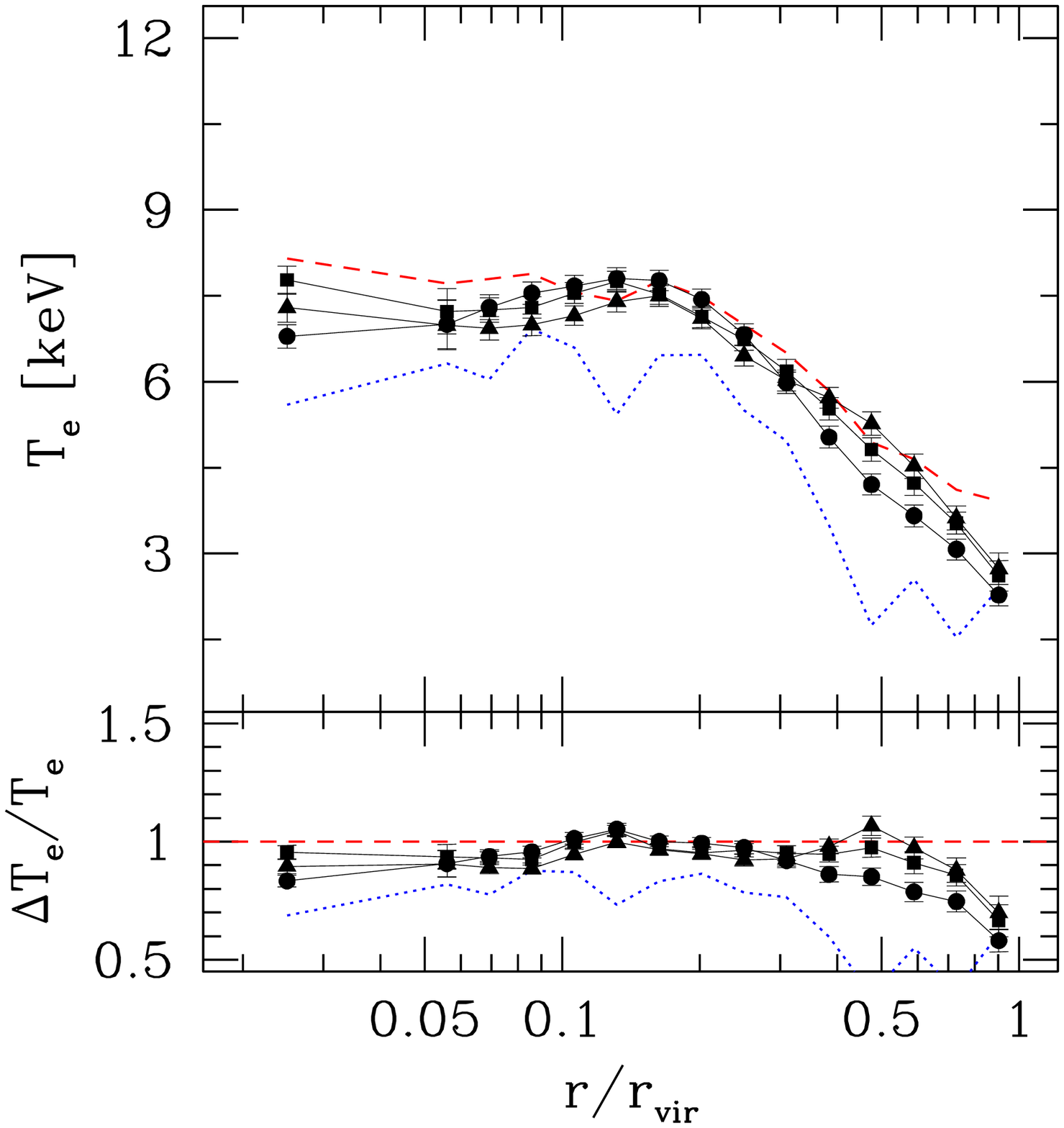,width=8.cm}
}
\caption{The same as in Figure \ref{fi:c9964jof.vir}, but for the C3
  cluster.}
\label{fi:c5726jof.vir}
\end{figure*}

\begin{figure*}
\centerline{ 
\psfig{file=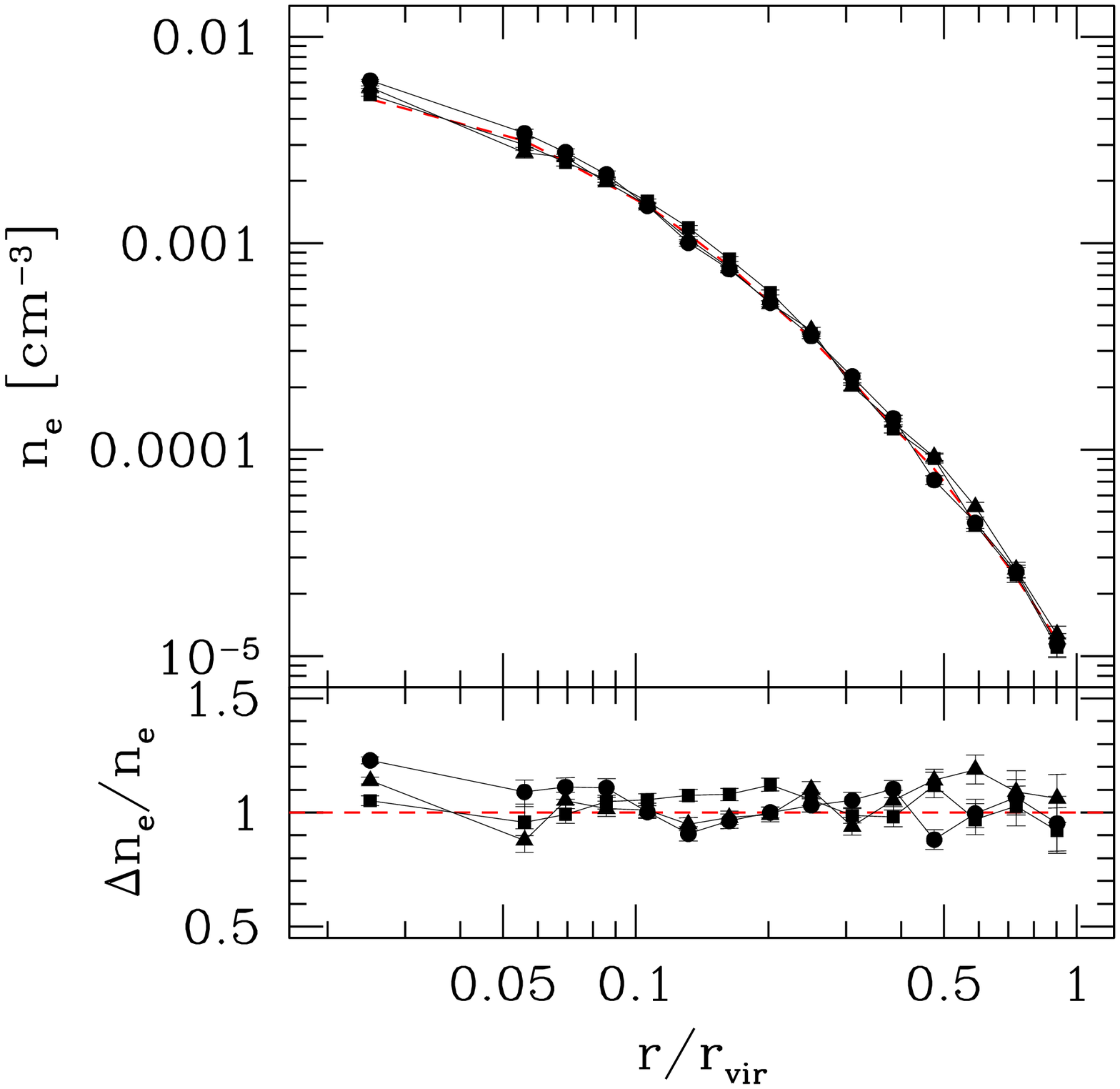,width=8.cm}
\psfig{file=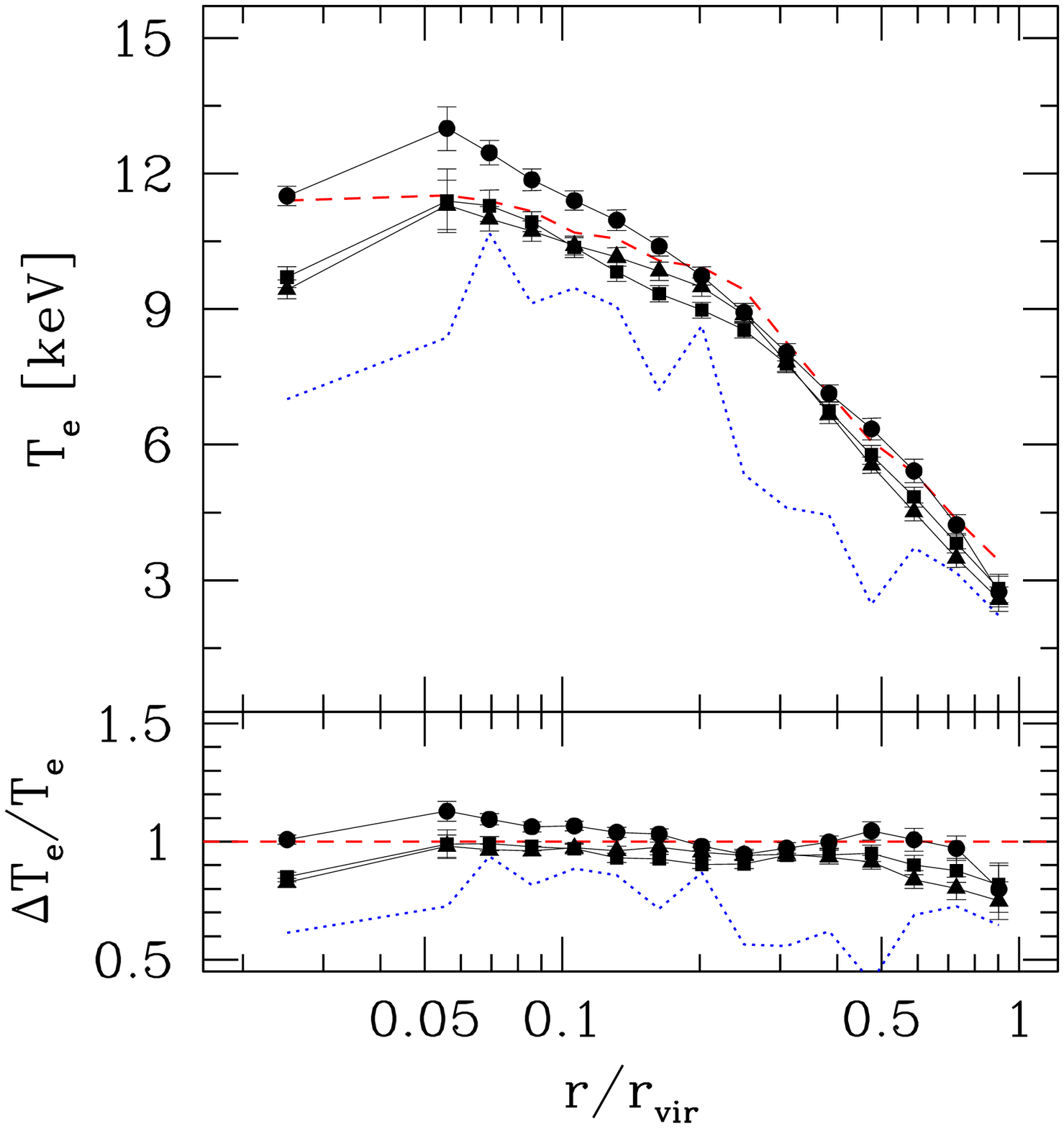,width=8.cm}
}
\caption{The same as in Figure \ref{fi:c9964jof.vir}, but for the C4
  cluster.}
\label{fi:c001jof.vir}
\end{figure*}

\subsection{tSZ and X--ray maps}\label{par:maps}

Around each cluster we extract a spherical region extending out to 6
$R_{vir}$.  Following \cite{2005MNRAS.356.1477D}, we create maps of
the relevant quantities along three orthogonal directions, extending
out to about 2 $R_{vir}$ from the cluster center, by using a
regular $512\times 512$ grid.

A number of different analyses, based on a joint deprojections of SZ
and X--ray cluster maps under the assumption of axial symmetry,
indicate that the X--ray selection tends to favor objects which are
elongated along the line-of-sight \citep[e.g., ][, and references
therein]{2005ApJ...625..108D}. In order to control the effect of this
selection bias, we decided to choose the axes of projection to be
aligned with the principal axes of inertia of the cluster. This will
allow us to quantify the difference in the reconstructed profiles when
the projection direction is that of maximum cluster elongation.

To derive these axes, we diagonalize the inertia tensor, which is given by
\be
I_{ij}=\sum_{p=0}^N (r_ir_j) 	\rho_p^2
\ee
where $i,j=0,1,2$ are the coordinate axes, $r_i$ is the $i$-th
coordinate of the particle $p$ with density $\rho_p$ and the sum is
extended over all the gas particles. We weight each particle by
$\rho_p^2$ so as to mimic the elongation in the X--ray emissivity.

The eigenvectors of the $I$ tensor provide the principal axes of the
best--fitting ellipsoid. The semi--axes $a_i$ of this ellipsoid are
proportional to square root of the corresponding eigenvalue $a_i
\propto \sqrt \lambda_i$ \citep[e.g.][]{1991MNRAS.249..662P}. We
choose the direction of projection $z$ to be that corresponding to
the largest semi--axis (i.e. the maximum elongation), while the $y$
and $x$ directions correspond to the medium and to the lower
semi--axes, respectively.

In the Tree+SPH code, each gas particle has a smoothing length $h_i$
and the thermodynamical quantities it carries are distributed within
the sphere of radius $h_i$ according to the compact kernel: \be W(x) =
\frac{8}{\pi h_i^3} \left\{
\begin{array}{ll}
1-6x^2+6x^3 & 0 \leq x \leq \frac{1}{2}\\
2(1-x)^3    & \frac{1}{2} \leq x \leq 1\\
0           & x \geq 1
\end{array}
\right.  \ee Here, it is $x = r/h_i$ with $r$ the distance from the
particle position. Therefore, we distribute the quantity of each
particle on the grid points within the circle of radius $h_i$ centered
on the particle.  Specifically, we compute a generic quantity $q_{jk}$
on the $(j,k)$ grid point as $q_{jk}d^2_p = \int q(r) dld^2_p = \sum
q_i (m_i/\rho_i )w_i$ where $d^2_p$ is the pixel area, the sum runs
over all the particles, and $w_i \propto\int W(x) dl$ is the weight
proportional to the fraction of the particle proper volume,
$m_i/\rho_i$, which contributes to the $(j,k)$ grid point. For each
particle, the weights $w_k$ are normalized to satisfy the relation
$\sum w_k = 1$ where the sum is over the grid points within the
particle circle.  When $h_i$ is so small that the circle contains no
grid point, the particle quantity is fully assigned to the closest
grid point.

As for the X--ray maps, they have been generated in the [0.5-2] keV
energy band, by computing the emissivity of each gas particle with a
Raymond-Smith code \citep{1977ApJS...35..419R}, assuming zero
metallicity in the cooling function.

Noise is finally added as described in Section \ref{par:noise}. We fix
the total number of photons in the virial radius to $10^4$ also for
simulated clusters. For the tSZ map, we adopt a noise level of 10
$\mu$K/beam for the objects having spectroscopic temperature $T_{sl}
>4$ K and 3 $\mu$K/beam for those having $3$ K $ < T_{sl} <4$ K.

\subsection{Results}

Having tested the reliability of the deprojection method, with the
regularization of the likelihood, we apply now this technique to the
more realistic case of hydrodynamical simulations. In this case, a
number of effects, such as deviations from spherical symmetry,
presence of substructures and presence of fore/background
contaminating structures, are expected to degrade the capability of
the deprojection to recover the three-dimensional profiles.

We show in detail the results on the density and temperature
deprojection the selected subset of four clusters presented in Table
\ref{tab:clusters} while the whole set of 14 clusters will be used to
assess on a statistical basis the efficiency with which the total gas
mass can be recovered.

The C2 and C4 objects are rather typical examples of our set of
clusters. They are fairly relaxed and with a modest amount of
substructures. As for the presence of substructures, they are well
known to represent an important source of bias in the deprojection,
especially of the X--ray signal, which is highly sensitive to gas
clumping. In order to remove this contaminating signal, we follow the
same method that is often adopted in the analysis of observational
data.  We first identify the detectable clumps by visual inspection of
the X--ray maps. The corresponding regions are then masked out both in
the X--ray and in the tSZ maps. The masked regions are excluded from
the computation of the signals to be deprojected. This leads to an
increase of the statistical uncertainties in those rings which have a
significant overlap with the masked regions. Clearly, due to the
finite photon statistics in the X--ray maps, small clumps may fall
below the detection threshold, while their presence may still affect
the emissivity.

 The recovered density and temperature profiles of C2 and C4 are shown
in Figures \ref{fi:c29931jof.vir} and \ref{fi:c001jof.vir}. Once all
the detectable clumps are masked out the reconstruction of the density
profile is generally good, but with a systematic overestimate of
$\sim$5 per cent, that we attribute to a residual small--scale gas
clumping. Although this effect is rather small, its presence
highlights the need to have a sufficient photon--count statistics to
identify gas inhomogeneities and remove their contribution in the
deprojection procedure.  The slight density overestimate corresponds,
as expected, to a small underestimate of the temperature, which is
forced by the requirement of reproducing the tSZ signal, $y\propto n_e
T_e$. For these two objects we also note that there are rather small
differences in the 3D profiles recovered from three orthogonal
projection directions, thus indicating that they are almost spherical
and without significant substructures along the different projection
directions. Errorbars are always of the order of a few percent, in
both density and temperature. We stress that these very small
errorbars, especially in temperature, are partly a consequence of the
regularization constraint.

As for the C3 cluster, we note that it has larger substructures
which will have a stronger impact on the recovered profiles.  Even
after masking all the detectable substructures, we still have a number
of unresolved clumps. As expected, in this case the density profile
(see Figure \ref{fi:c5726jof.vir}) is overestimated by a larger
factor, $\sim 10$ cent, with a corresponding more significant
underestimate of the temperature. The deviations of the deprojected
profiles in the outer parts are also larger. This is due to stronger
contaminations from the fore/background structures, which are both
placed at the outskirts of the cluster and along the projection
direction, in the cosmic web surrounding the cluster.  In fact, both
tSZ and X--ray maps are produced by projecting a region of 6 $R_{vir}$
in front and in the back of the cluster center.

\begin{figure}
\centerline{ 
\psfig{file=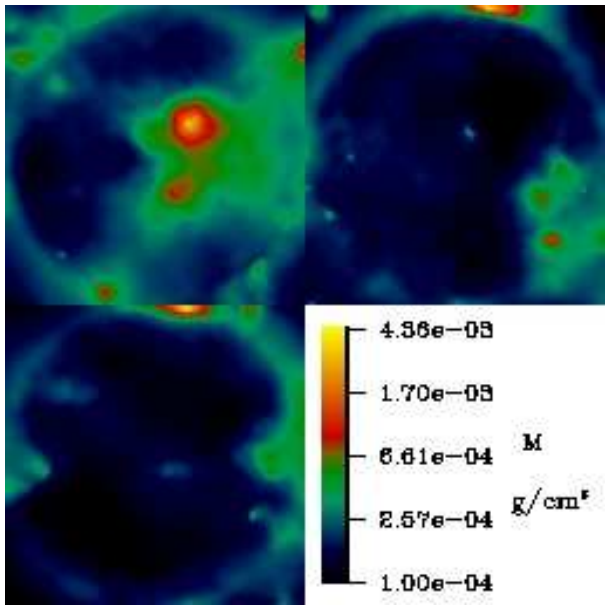,width=8cm}
}
\caption{Projected surface gas mass density of the C1 cluster along
  the $z$ (top left), $y$ (top right) and $x$ (bottom left)
  directions, after removing the contribution from all gas particles
  which inside the virial region of the cluster.}\label{fi:c9964.bkg}
\end{figure}
In this respect, the C1 cluster is particularly interesting.  Along
the $x$-axis projection there is a merging groups along the same line
of sight, at a distance of $\sim 1.2 R_{vir}$ from the center of the
main cluster. In Figure \ref{fi:c9964.bkg} we show the projected mass
surface density of the gas along the three projection directions,
after removing the mass of the main cluster within $R_{vir}$.  While
the residual mass surface density is quite small along the $x$ and $y$
directions, a presence of a gas clump are shows up in the $z$
projection. While this structure provides a rather small contribution
to the X--ray signal, its gas pressure is comparable to that of the
main cluster, thereby significantly contaminating the tSZ effect
signal.  As a consequence, the density profile (see Figure
\ref{fi:c9964jof.vir}) is essentially unaffected, while the
temperature is clearly boosted by $\sim 20$ per cent with respect to
the that obtained from the other two projections. Although this is a
quite peculiar case, in which the secondary structure is relatively
large and aligned with the main cluster along the line of sight, it
illustrates the role of projection contamination from unidentified
structures in recovering the 3D thermal structure of the ICM.

\begin{figure*}
\centerline{ 
\psfig{file=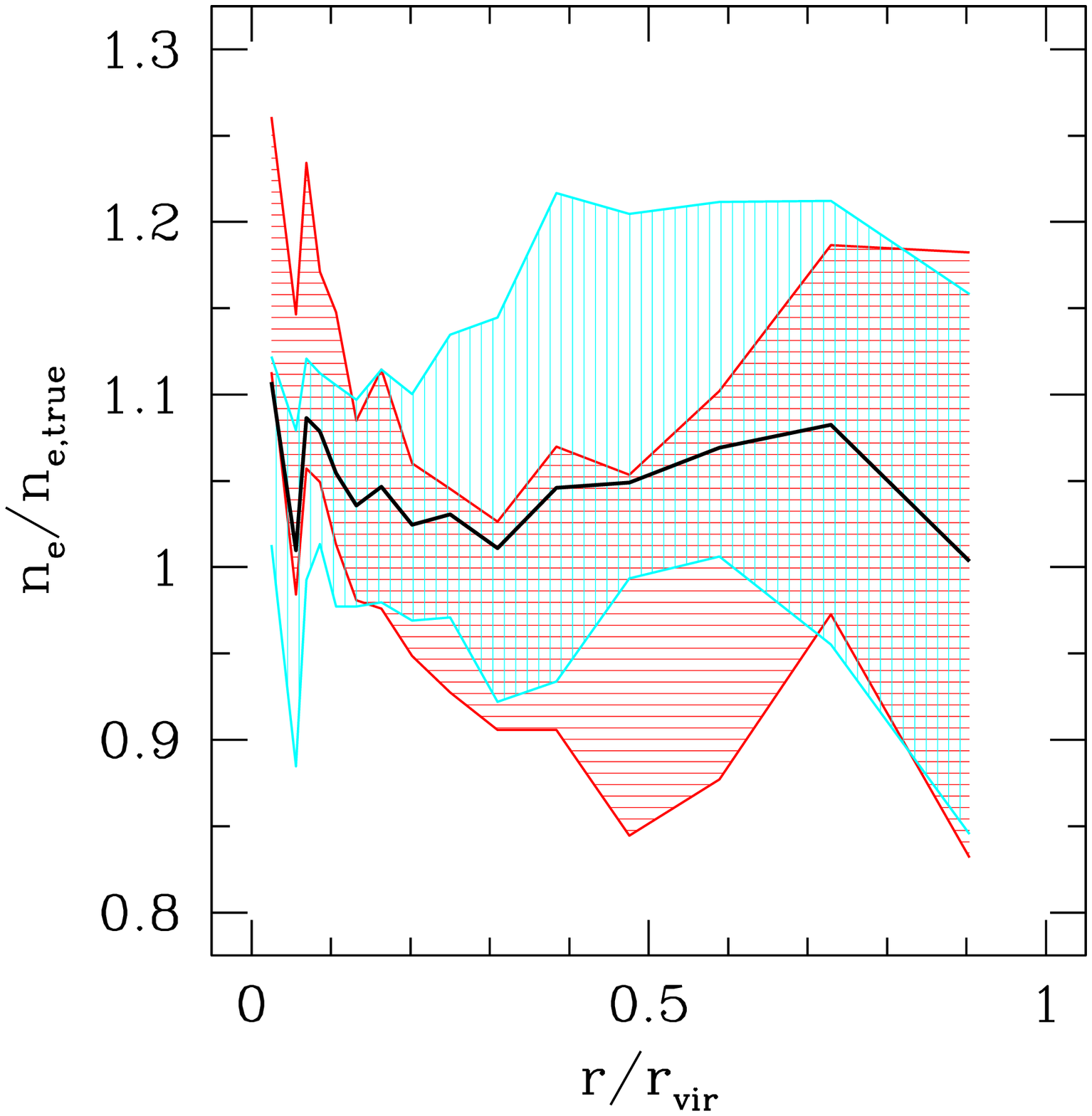,width=8cm}
\psfig{file=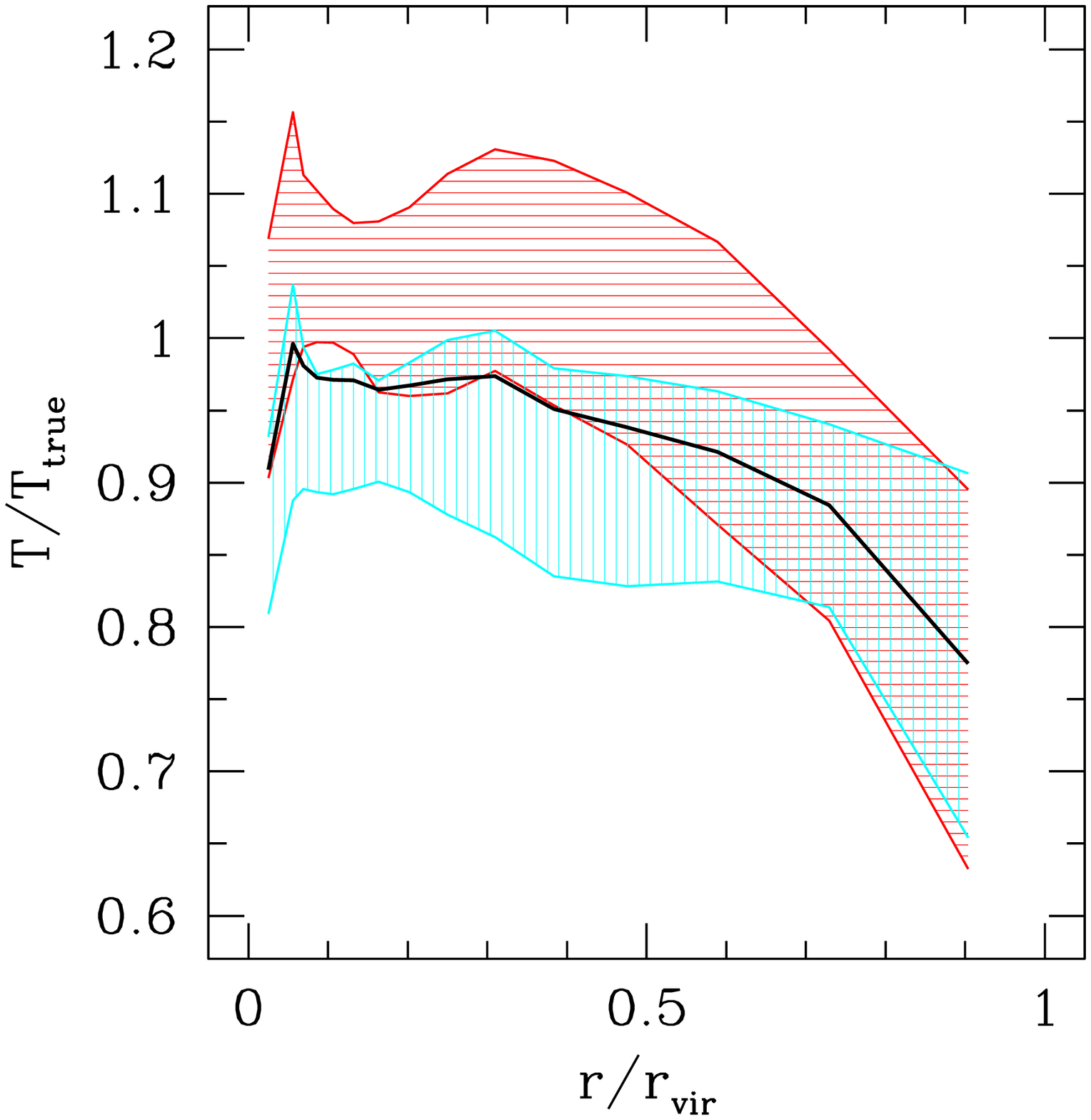,width=8cm} 
}
\caption{The ratio between the reconstructed and the true
(mass--weighted) temperature and density profiles. The shaded areas
encompass the $\pm 1\sigma$ regions of the recovered profiles over the
ensemble of simulated clusters. The
horizontally shaded area is for the projections along the $z$ axis,
while the vertically shaded area is for the projections along the other two
axes. The black line shows the mean over all projections of all
clusters.}
\label{fi:temp_scatter}
\end{figure*}

We note in Figs.\ref{fi:c9964jof.vir}--\ref{fi:c001jof.vir} that the
density profiles recovered from the projection of maximum elongation
are overestimated at small radii, while they are underestimated in the
outskirts. In order to quantify this effect, we show in the left panel
of Figure \ref{fi:temp_scatter} the ratio between the true and the
reconstructed density profiles, after averaging over the sample of
simulated clusters. By averaging over all the projection directions,
the density is generally overestimated by about 5 per cent at all
cluster radii. This result is confirmed also by analyses performed on
synthetic X--ray observations of simulated clusters
\citep[][]{2006MNRAS.369.2013R, 2007ApJ...655...98N}. On the other
hand, the density profile reconstructed from the projection along the
$z$ axis is confirmed to be significantly larger than along the other
direction in the very inner part, with an inversion at $r\magcir
0.2r_{\rm vir}$. Indeed, the elongation causes the objects to appear
more compact in the X--ray maps, which drive the density
reconstruction. This boosts the deprojected central density, while
depletes it in the outskirts.

As shown in the right panel of Fig. \ref{fi:temp_scatter}, the
temperature is generally underestimated by \mincir 10 per cent out to
$\simeq 0.7 r_{vir}$. At larger radii this underestimate increases,
reaching a mean value of about 20 per cent at $r_{vir}$, as a
consequence of the relatively larger contamination by fore/background
structures. The scatter is generally larger than the uncertainty
introduced by the noise, so that it has to be considered as intrinsic
to the measure. This scatter has different origins, such as unresolved
gas clumps, asphericity of the clusters, fore--background
contaminations.  In general, the temperature recovered from the
projection along the $z$ axis is slightly larger than the one from the
other two axes. The difference is more apparent in the central regions
and becomes smaller in the outskirts. The reason for this behaviour is
that the temperature reconstruction is more affected by the tSZ
signal. Along the direction of maximum elongation, this signal in
enhanced since the ICM pressure is integrated along a larger path. The
tSZ signal receives then an important contribution from cluster
regions where the density is underestimated. As a result the
reconstructed temperature is correspondingly increased to compensate
for this effect.  

In Figs. \ref{fi:c9964jof.vir}--\ref{fi:c001jof.vir} we also show the
three--dimensional profile of the spectroscopic--like temperature
(dotted line), which has been shown by \cite{2004MNRAS.354...10M} to
represent a quite close proxy to the actual X--ray temperature
obtained from a spectroscopic fit \citep[see
also][]{2006ApJ...640..710V}. This temperature is computed as
\be
T_{sl}\,=\,{\sum_i \rho_i m_i T_i^{\alpha -1/2}\over \sum_i \rho_i m_i
  T_i^{\alpha -3/2}}\,, 
\label{eq:tsl}
\ee
with $\alpha \simeq 0.75$ and the sum extends over all the gas
particles having internal energy larger than 0.5 keV. According to its
definition, this temperature gives more weight to the low--temperature
phase in a thermally complex ICM. This is the reason for the drop of
the $T_{sl}$ profile at the cluster center and for the wiggles which
mark the positions of merging sub-clumps which are relatively colder
than the ambient ICM. In general, the profile of $T_{sl}$ are lower
than those of the electron temperature, by an amount which is larger
for hotter systems \citep[see also][]{2005ApJ...618L...1R}. These
figures highlight that the temperature profiles, as obtained from our
deprojection analysis, are much closer to the mass--weighted
temperature, which measures the total thermal content of the ICM, than
to $T_{sl}$. An important consequence of this difference will clearly
be the estimate of the total cluster mass from the application of
hydrostatic equilibrium. We will discuss the application of our
deprojection method to cluster mass estimates in a forthcoming paper.

\begin{figure}
\centerline{
\hbox{
\psfig{file=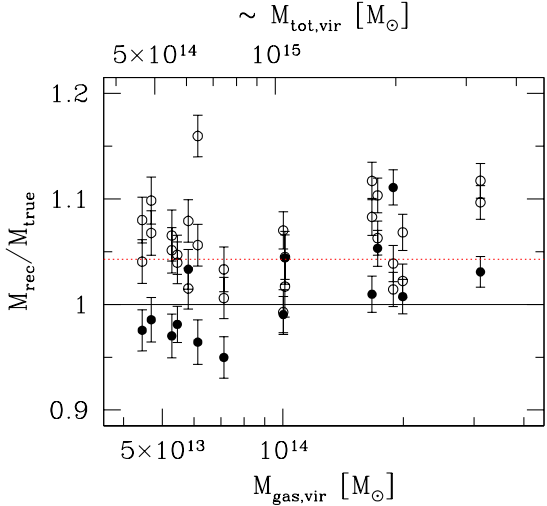,width=8.cm}
}}
\caption{The ratio between the recovered and the true values of
  the total gas mass for simulated clusters out to For each cluster we
  show the result of the deprojection along the three orthogonal
  directions, with the projection corresponding to the maximum
  elongation being marked with a filled circle. Errorbars correspond
  to the $1\sigma$ confidence level, by accounting for the full error
  correlation matrix when integrating the 3D gas density profiles. The
  horizontal dotted line shows the average value of the ratio.}
\label{fi:jof_mass}
\end{figure}

\begin{figure*}
\centerline{
\hbox{
\psfig{file=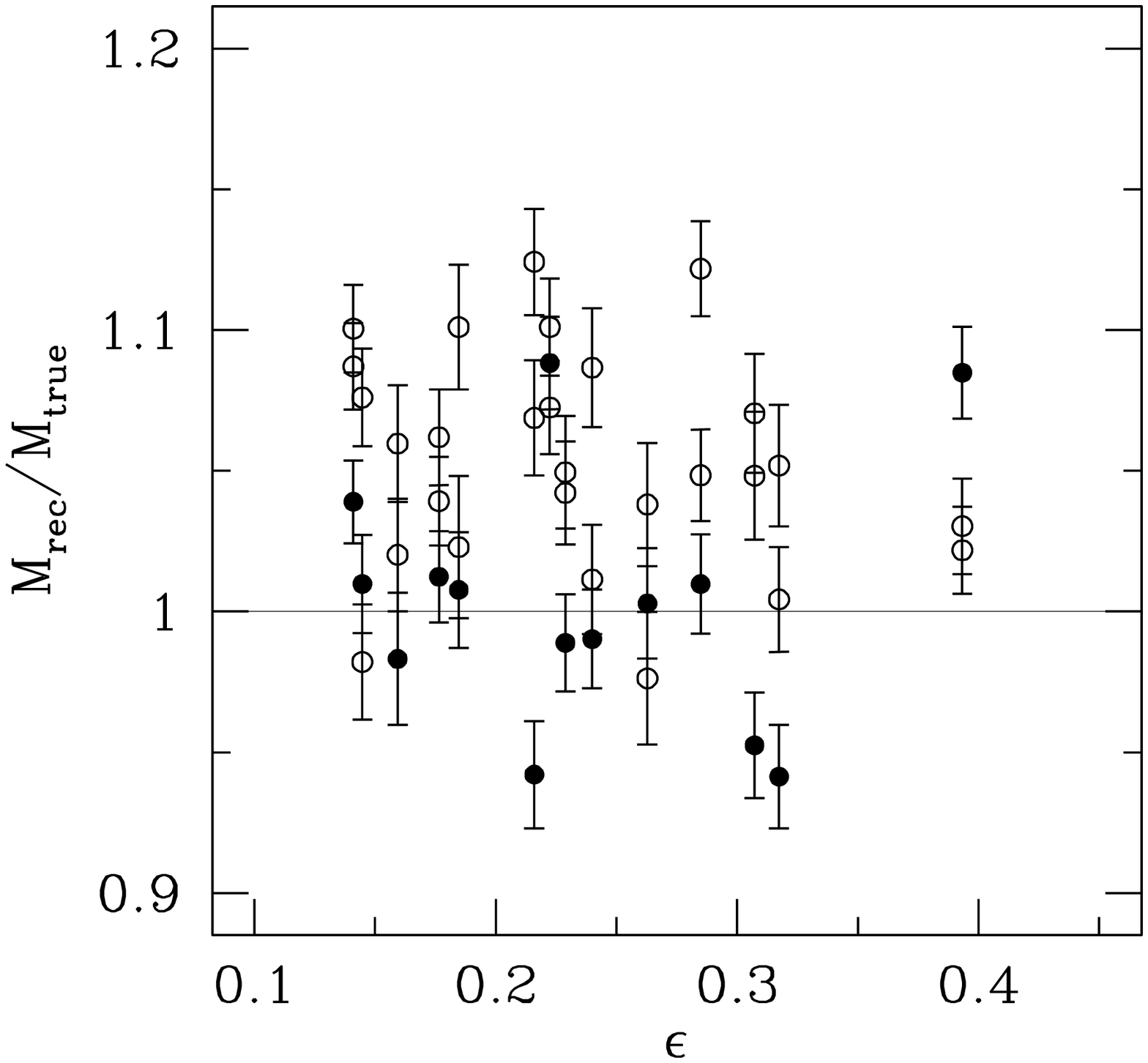,width=6.cm}
\psfig{file=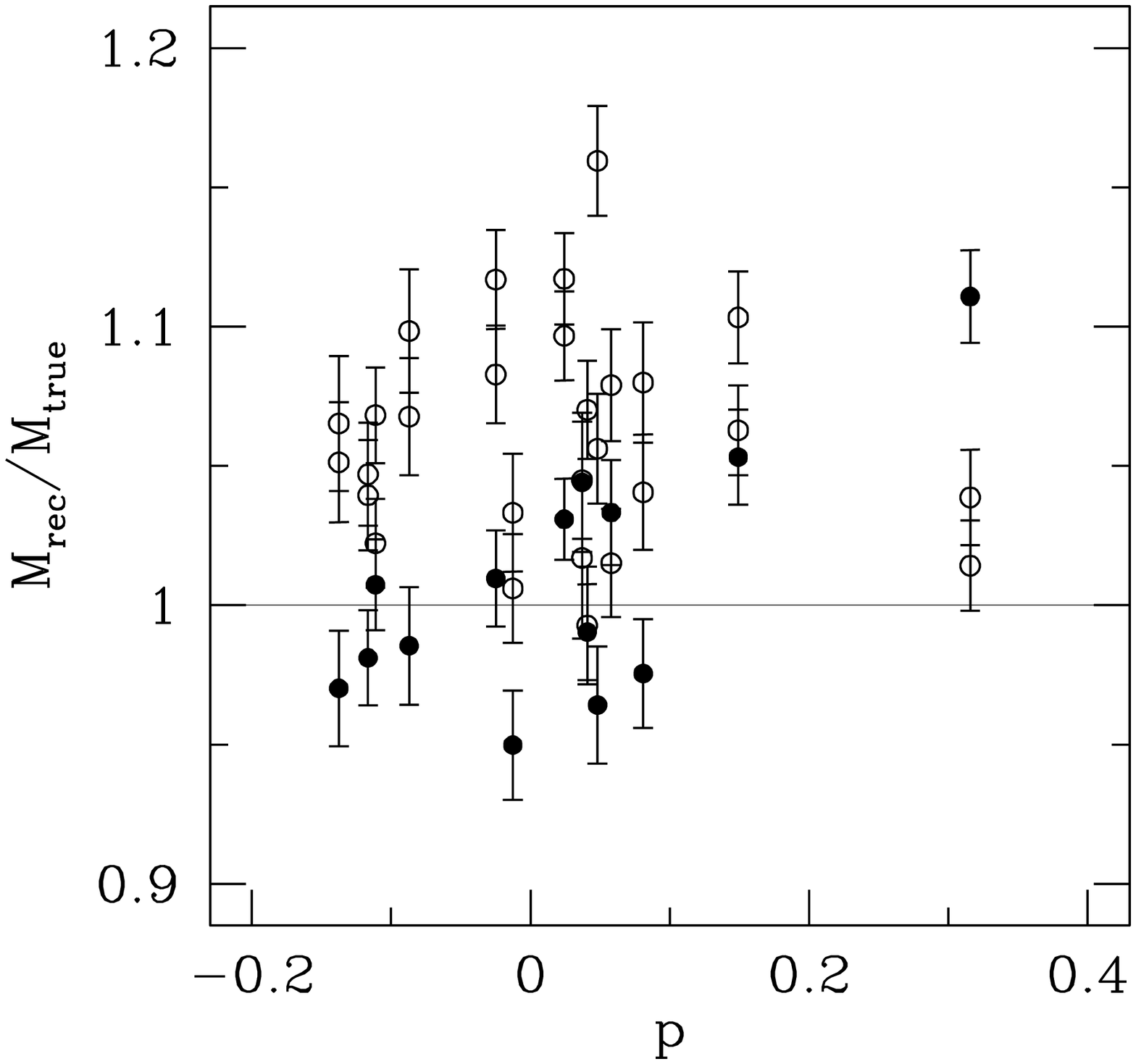,width=6.cm}
\psfig{file=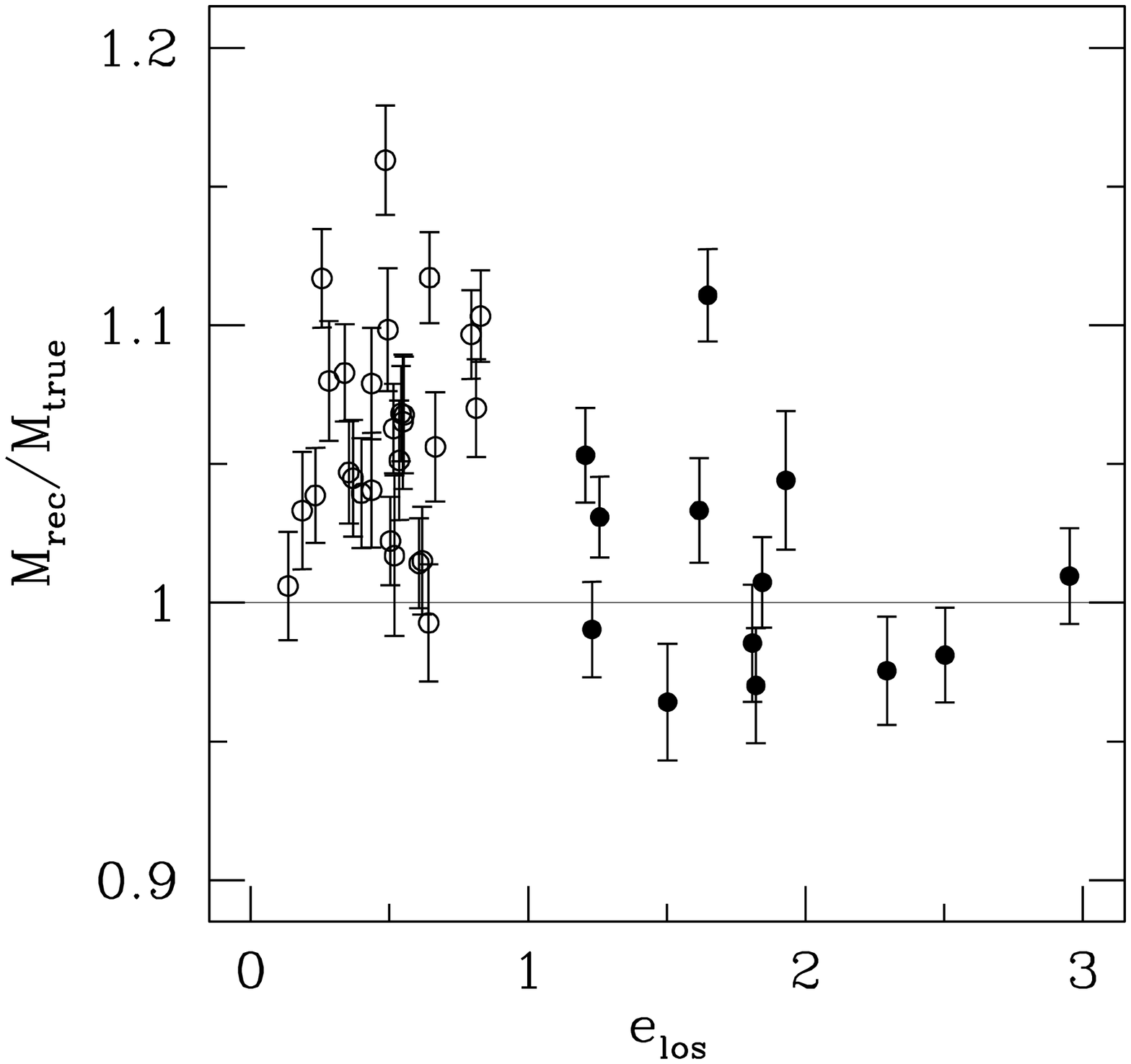,width=6.cm}
}}
\caption{The same of figure \ref{fi:jof_mass}, but as a function
  of cluster ellipticity (left panel), prolateness (central panel) and
  elongation along the line of sight (right panel).}
\label{fi:mass_ellipt}
\end{figure*}

\subsection{Recovering the gas mass}
As a first application of our deprojection procedure, we compute the
gas mass of the clusters, which is calculated simply by summing up the
mass contained into each radial bin. Since the bins are not
independent, the errors on the total gas mass have been calculated by
using both variances and covariances of the values of the density at
different radii.
\be
\sigma_M^2= {1 \over N-1} \sum_{i=0}^N \sum_{j=0}^N \sigma_{m,ij},
\ee
where $\sigma_{m,ij}$ is the covariance between the mass content
of the $i$-th and of the $j$-th shells, directly obtained from the
covariance between the gas density in different bins (e.g., see left
panel of Fig. \ref{fi:beta_corr}). Note that since the covariance
between the density in adjacent bins is generally negative, neglecting
it would lead to a systematic overestimate of the error on the mass.

We give the results on the estimate of gas mass for the whole set of
14 simulated clusters.  The small overestimate found for the density
profiles is obviously propagated to the estimate of the total gas
mass. The resulting bias turns out to be very small, {and amounts to
about 4 percent}, with no obvious trend with the cluster mass.  This
demonstrate that residual gas clumping, after the removal of the
substructures identified in the X--ray maps, has a small effect on the
our capability of the recovering the total mass of the ICM. We note
that cluster-by-cluster variance is often comparable to the
``projection variance'', i.e. to the differences found when projecting
the same clusters along different directions. We also note that the
uncertainties in the individual $M_{gas}$ estimates, typically of the
order of a few per cent, are smaller than the scatter. This indicates
that the intrinsic scatter in the recovered gas mass is in fact
associated to the deviations of the simulated clusters from perfect
spherical symmetry.

\subsection{The effect of morphology}
In order to better understand how morphology affects the deprojection,
we plot in Figure \ref{fi:mass_ellipt} the recovered gas mass as a
function of cluster ellipticity, prolateness and elongation along the
line of sight.  The ellipticity of a triaxial object is defined
as:
\be
\epsilon={1 \over 2} {{1\over a_{min}} -{1\over a_{max}}
\over {1\over a_{min}} +{1\over a_{med}}
+{1\over a_{max}}}
\ee
and the prolateness as:
\be
p={1 \over 2} {{1\over a_{min}} -{2\over a_{med}}
+{1\over a_{max}} \over {1\over a_{min}}
+{1\over a_{med}} +{1\over a_{max}}}
\ee
The elongation is defined as the ratio between the semi--axis aligned
with the line of sight and the larger of the other two semi--axes.

The gas mass recovered from the projection along the principal axis is
generally lower than those from the other two projections.  This
underestimate generally is anticorrelated with the elongation of the
cluster, although with a substantial scatter.  From the left and the
central panels of Figure \ref{fi:mass_ellipt}, we do not find any
significant correlation between the global 3D morphology of the
clusters (prolateness and ellipticity) and the bias in the
deprojection. Instead, as shown in the right panel, any effect in the
gas mass recovery is driven by the orientation of the cluster.

\section{Discussion and conclusions}
We have presented results of deprojection methods, aimed at recovering
the three-dimensional density and temperature profiles of galaxy
clusters, by combining X--ray surface brightness and thermal SZ (tSZ)
maps. The main aim of our analysis is to verify to what accuracy one
can recover the thermal structure of the ICM by taking advantage of
the different dependence of the X--ray and tSZ signal on the gas
density and temperature, thereby avoiding performing X--ray
spectroscopy. The two deprojection methods considered are both based
on assuming spherical symmetry of the clusters.

The first one follows a geometrical approach, in which the 3D profiles
are recovered with an iterative procedure that deprojects the observed
images starting from the outermost ring and proceeding inwards.  The
second method assumes the values of the 3D gas density and temperature
profiles at different radii and computes from them the expected SZ and
X--ray surface brightness which is then compared to the observations
with a maximum likelihood approach.  In the computation of the
likelihood, we also introduced a regularization term, which allows us
to suppress spurious oscillations in the recovered profiles. Using a
Monte Carlo Markov Chain (MCMC) approach to optimize the sampling of
the parameter space, this second method also allows us to recover the
full correlation matrix of the errors in the parameter fitting.

The main results of our analysis can be summarized as follows.
\begin{itemize}
\item The application of both methods to an idealized spherical
  polytropic $\beta$--model shows that the 3D profiles are always
  recovered with excellent precision (of about 3--4 per cent), thus
  demonstrating that such methods do not suffer from any intrinsic
  bias.
\item The application of the maximum--likelihood method to
  hydrodynamical simulations of galaxy clusters always provides
  deprojected profiles of gas density and temperature, which are in
  good agreement with the true ones, out to the virial radius. We find
  a small ($\mincir 10$ per cent) systematic overestimate of the gas
  density, which is due to the presence of some residual gas clumping,
  which is not removed by masking out the obvious substructures
  identified in the X--ray maps.
\item The total gas mass is recovered with a small bias of 4 per cent,
  with a sizable scatter of about 5 per cent. This result shows that
  residual gas clumping should have a minor impact in the estimate of
  the total gas mass. We do not find any trend in the recovery of the
  gas mass with the total cluster mass.
\item The gas mass reconstructed along the maximum elongation axis is
  generally lower (by up to 10 per cent) with respect to the mass
  reconstructed along the other two projection axes, the size of this
  effect being larger for more elongated clusters.
\item The temperature is generally well recovered, with $\sim 10$ per
  cent deviations from the true one out to $\simeq 0.7 R_{\rm
    vir}$. The rather small size of this bias confirms that the
  combination with tSZ data is a valid alternative to X--ray
  spectroscopy for temperature measurements. The temperature
  reconstructed from the projection along the axis with maximum
  elongation is slightly higher than those from the other two axes,
  particularly in the inner regions.
\end{itemize}

Our results confirm the great potentials of combining spatially
resolved tSZ and X--ray observations to recover the thermal structure
of the ICM. This approach has several advantages with respect to the
traditional one based on X--ray spectroscopy. First, the temperature
recovered from the fit of the X--ray spectra is known to provide a
biased estimate of the total thermal content of the ICM, the size of
this bias increasing with the complexity of the plasma thermal
structure \citep[e.g.,
][]{2004MNRAS.354...10M,2006ApJ...640..710V}. Secondly, X--ray surface
brightness profiles can be obtained with good precision with a
relatively small number of photon counts. Also, once the cosmic and
instrumental backgrounds are under control, the surface brightness can
be recovered over a large portion of the cluster virial regions, as
already demonstrated with ROSAT-PSPC imaging data
\citep[e.g.,][]{1999ApJ...525...47V,2005A&A...439..465N}. Since the
tSZ has the potential of covering a large range in gas density, then
its combination with low--background X--ray imaging data will allow
one to recover the temperature profiles out to the cluster's
outskirts.

A limitation of the analysis presented in this paper is that we did
not include realistic backgrounds in the generation of the X--ray and
tSZ maps. As we have just mentioned, there are reasonable perspectives
for a good characterization of the X--ray background. However, the
situation may be more complicated for the tSZ background. In this
case, contaminating signals from unresolved point--like radio sources
\citep[e.g.,][]{2006A&A...447..405B} and fore/background galaxy groups
\citep[e.g.,][]{2007arXiv0704.2607H} could affect the tSZ signal in
the cluster outskirts. In this respect, the possibility of performing
multi--frequency observations with good angular resolution will surely
help in characterizing and removing these contaminations.

Single--dish sub-millimetric telescopes of the next generation
promises to provide tSZ images of clusters with a spatial resolution
of few tens of arcsec, while covering fairly large field of views,
with 10--20 arcmin aside, with a good sensitivity. At the same time,
planned satellites for X--ray surveys\footnote{eROSITA:
\tt{http://www.mpe.mpg.de/projects.html\#erosita}}\footnote{EDGE: {\tt
http://ibis.rm.iasf.cnr.it/EdgeOverview.htm}} will have the capability
of surveying large areas of the sky with a good quality imaging and
control of the background. These observational facilities will open
the possibility of carrying out in survey mode high--quality tSZ and
X--ray imaging for a large number of clusters. The application of
deprojection methods, like those presented in this paper will provide
reliable determinations of the temperature profiles and, therefore, to
exploit their potentiality as tools for precision cosmology.

\section*{Acknowledgments.}
We would like to thank Stefano Ettori for his help with the
geometrical deprojection algorithm. We also thank Antonaldo Diaferio,
Sunil Golwala, Silvano Molendi, Manolis Plionis, Elena Rasia and Paolo
Tozzi for useful discussions. This work has been partially supported
by the PD-51 INFN grant and by the ASI-INAF I/023/05/0 grant. E.P. is
an ADVANCE fellow (NSF grant AST-0649899).

\bibliographystyle{mn2e}

\bibliography{master}

\clearpage

\end{document}